\documentclass[12pt]{article}

\usepackage{graphicx}
\usepackage{epsf,amsfonts,hyperref}
\renewcommand{\appendix}[1]{
    \addtocounter{section}{1}
    \setcounter{equation}{0}
    \renewcommand{\thesection}{\Alph{section}}
    \section*{Appendix \thesection\protect\indent #1}
    \addcontentsline{toc}{section}{Appendix \thesection\ \ \ #1}
}
\newcommand\encadremath[1]{\vbox{\hrule\hbox{\vrule\kern8pt
\vbox{\kern8pt \hbox{$\displaystyle #1$}\kern8pt}
\kern8pt\vrule}\hrule}}
\def\enca#1{\vbox{\hrule\hbox{
\vrule\kern8pt\vbox{\kern8pt \hbox{$\displaystyle #1$} \kern8pt}
\kern8pt\vrule}\hrule}}

\newcommand\figureframex[3]{
\begin{figure}[bth]
\hrule\hbox{\vrule\kern8pt \vbox{\kern8pt \vbox{
\begin{center}
{\mbox{\epsfxsize=#1.truecm\epsfbox{#2}}}
\end{center}
\caption{#3} }\kern8pt} \kern8pt\vrule}\hrule
\end{figure}
}
\newcommand\figureframey[3]{
\begin{figure}[bth]
\hrule\hbox{\vrule\kern8pt \vbox{\kern8pt \vbox{
\begin{center}
{\mbox{\epsfysize=#1.truecm\epsfbox{#2}}}
\end{center}
\caption{#3} }\kern8pt} \kern8pt\vrule}\hrule
\end{figure}
}

\renewcommand{\thesection}{\arabic{section}}

\makeatletter \@addtoreset{equation}{section} \makeatother
\newtheorem{theorem}{Theorem}[section]

\newtheorem{remark}{Remark}[section]
\newtheorem{proposition}{Proposition}[section]
\newtheorem{lemma}{Lemma}[section]
\newtheorem{corollary}{Corollary}[section]
\newtheorem{definition}{Definition}[section]
\def\br{\begin{remark}\rm\small}
\def\er{\end{remark}}
\def\bt{\begin{theorem}}
\def\et{\end{theorem}}
\def\bd{\begin{definition}}
\def\ed{\end{definition}}
\def\bp{\begin{proposition}}
\def\ep{\end{proposition}}
\def\bl{\begin{lemma}}
\def\el{\end{lemma}}
\def\bc{\begin{corollary}}
\def\ec{\end{corollary}}
\def\beaq{\begin{eqnarray}}
\def\eeaq{\end{eqnarray}}

\newcommand{\eq}[1]{Eq.~(\ref{#1})}

\newcommand{\beq}{\begin{equation}}
\newcommand{\eeq}{\end{equation}}
\newcommand{\bea}{\begin{eqnarray}}
\newcommand{\eea}{\end{eqnarray}}

\renewcommand{\and}{{\qquad {\rm and} \qquad}}

 \newcommand{\Tr}{{\,\rm Tr}\:}

\newcommand{\Res}{\mathop{\,\rm Res\,}}

\newcommand{\td}[1]{{\tilde{#1}}}

\newcommand{\ee}[1]{{{\rm e}^{#1}}}

\newcommand{\Pint}{{\int\kern -1.em -\kern-.25em}}

\newcommand{\ovl}{\overline}

\newcommand{\CP}{{\cal P}}

\begin{document}
%=============================Page de titre==============%\date{??}
%\author{Eynard}
%\title{Mixed correlation functions for hermitian random matrices}
%\topmargin .5cm \textheight 21.5cm \textwidth 15.8cm
%\oddsidemargin 0.54cm
%\evensidemargin 0.54cm
\sloppy

%\maketitle

\pagestyle{empty} \hfill SPhT-T07/118
\addtolength{\baselineskip}{0.20\baselineskip}
\begin{center}
\vspace{26pt} {\large \bf {Topological expansion and boundary conditions}}
\newline
\vspace{26pt}

{\sl B.\ Eynard}\hspace*{0.05cm}\footnote{ E-mail: eynard@spht.saclay.cea.fr }, {\sl N.\ Orantin}\hspace*{0.05cm}\footnote{ E-mail: orantin@spht.saclay.cea.fr }\\
\vspace{6pt}
Service de Physique Th\'{e}orique de Saclay,\\
F-91191 Gif-sur-Yvette Cedex, France.\\
\end{center}

\vspace{20pt}
\begin{center}
{\bf Abstract}:

\bigskip

In this article, we compute the topological expansion of all possible mixed-traces in a hermitian two matrix model.
In other words we give a recipe to compute the number of discrete surfaces of given genus, carrying an Ising model, and with all possible given boundary conditions.
The method is recursive, and amounts to recursively cutting surfaces along interfaces.
The result is best represented in a diagrammatic way, and is thus rather simple to use.

\end{center}

%-----------------------------ABSTRACT--------------------------------------
%
%Abstract

%\begin{center}
%In this short note, we extend the diagrammatic technic developped in \cite{} to compute the whole topological expansion
%of any correlation function of the formal hermitian two matrix model. That is to say, we compute the generating function
%of surfaces with given genus, number of boundaries and boundary condition on it.

%\end{center}

%\newpage
%\pagestyle{empty}

%\section*{}

%\newpage
\vspace{26pt} \pagestyle{plain} \setcounter{page}{1}

%*********************************************************************
%==================== ARTICLE =======================================%*********************************************************************

\section{Introduction}

\subsection{Counting surfaces with given boundary conditions}

The problem of boundary conditions is a very important one in statistical mechanics, conformal field theory, string theory... (see for example \cite{kostov, schomerus, ribault} for recent developments).
In this article we address the problem of counting configurations of an Ising model on a random lattice, with given boundary conditions. This problem can be equivalently stated as computing mixed traces expectation values in a 2-matrix model.

The 2-matrix model was introduced by Kazakov \cite{kazakovising} as the Ising model on a random lattice. Its partition function reads:
\beq
Z = \int dM_1 \, dM_2 \,\,\, \ee{-N\Tr[V_1(M_1)+V_2(M_2)-M_1 M_2]}
\eeq
where $V_1$ and $V_2$ are polynomials,
and where the integral is a {\bf formal} hermitian matrix integral (see for example \cite{eynform} for a definition of formal integrals), i.e. it is computed by first expanding the exponential of the non-quadratic part of $V_1$ and $V_2$, and then exchanging the sums and integrals. A formal integral is thus a formal series whose general terms are moments of gaussian integrals \cite{eynform}.

It is well known from Wick's theorem that such a formal integral is a combinatorial generating function which enumerates discrete surfaces (also called maps in the combinatorists litterature) whose faces can have 2 possible colors $1$ or $2$, or let us say $+$ or $-$, or blue or red.

\medskip

\begin{figure}
$$ \left<\Tr M_1^3 \Tr M_2^5\right>_c:=
\begin{array}{r}
{\epsfxsize 6cm\epsffile{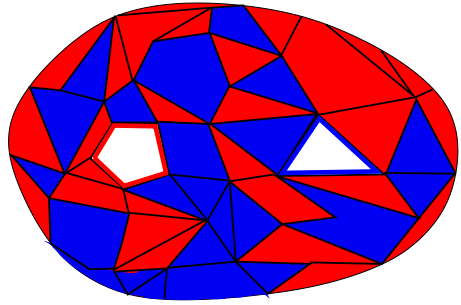}}
\end{array}$$
  \caption{Example of surface generated by $\left<\Tr M_1^3 \Tr M_2^5\right>_c$, where one has associated the
  blue color to $M_1$ and the red color to $M_2$: it is a cylinder with one boundary of length 5 with red condition and one boundary of length 3 with blue condition.}\label{fig1}
\end{figure}

The moments:
\beq
<\Tr M_1^{l}>
\eeq
are generating functions for discrete connected surfaces with one boundary of color $1$ and length $l$ (more precisely, surfaces with one marked face of color $1$ and of degree $l$, and one marked edge on the boundary, removed from a closed surface).
Similarly, $<\Tr M_2^{l}>$  is a generating function which counts surfaces with one boundary of color $2$ and length $l$.
More generally,
$<\Tr M_1^{l_1} \Tr M_1^{l_2} \dots \Tr M_1^{l_m} \Tr M_2^{l'_1} \Tr M_2^{l'_2} \dots \Tr M_2^{l'_{m'}} >_c$ is a generating function which counts connected surfaces with $m$ boundaries of color $1$ and respective lengths $l_1,\dots,l_m$, and $m'$ boundaries of color $2$ and respective lengths $l'_1,\dots,l'_{m'}$ (see fig.\ref{fig1} for an example). The subscript $<>_c$ in the expectation values means "connected part" or "cumulant", it ensures that only connected surfaces appear in the Wick expansion.

\begin{figure}
$$ \left<\Tr M_1^2 M_2^5\right>:=
\begin{array}{r}
{\epsfxsize 6cm\epsffile{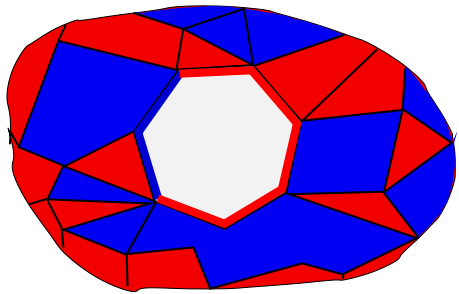}}
\end{array}$$
  \caption{Example of surface generated by $\left<\Tr M_1^2 M_2^5\right>_c$: it is a disc with one boundary of length 7 with
   red condition for 5 adjacent segments followed by two segments with blue condition.}\label{fig2}
\end{figure}

More interesting is:
\beq
< \Tr M_1^{l} M_2^{l'} >.
\eeq
It is a generating function which counts surfaces with only one boundary of length $l+l'$, with $l$ color $1$ sites followed by $l'$ color $2$ sites (see fig.\ref{fig2} for an example).

And more generally,
\beq
< \Tr M_1^{l_1} M_2^{l'_1}  M_1^{l_2} M_2^{l'_2} \dots >
\eeq
counts surfaces with one boundary of length $\sum l_i+ l'_i$ with $l_1$ sites of color $1$ followed by $l'_1$ sites of color $2$ then $l_2$ sites of color $1$, ..., etc.

\medskip
It is easy to see that one can design such observables for any given boundary conditions: any number of boundaries, and any pattern of sites on the boundaries.

\bigskip
In this article we show how to compute those generating functions for surfaces of given topology.

\subsection{Outline and main results}

The paper is organized as follows:

\begin{itemize}

\item in section 2, we summarize briefly some previous knowledge of formal 2-matrix model integrals.
Namely, we recall how to compute the "disc amplitude", and the spectral curve, and from there the result of \cite{EOFg}, i.e. how to count surfaces with uniform boundary conditions.

\item in section 3, we define appropriate notations for describing arbitrary boundary conditions.
We recall which cases were already known in the literature.

\item in section 4, we give the formula for computing the generating functions counting surfaces of any genus and arbitrary boundary conditions.
The formula is best represented diagrammatically, and has a very intuitive interpretation.

\item in section 5, we show some examples of applications of our formula, and in particular we show how to recover previously known cases.

\item section 6 is the conclusion.

\item the proof of the main formula of section 4, is written in the appendix, because it is rather technical.

\end{itemize}

\section{Reminder 2-matrix model}

The 2-matrix model has generated a considerable number of works.
Here, we use the method of loop equations \cite{Migdal, ZJDFG, staudacher}, which is well suited for genus expansion computations.

\subsection{The resolvent}

The resolvent is defined as:
\beq
\ovl{W}_1(x) = \left<\Tr {1\over x-M_1}\right> = \sum_{l=0}^\infty {1\over x^{l+1}}\,\, \left<\Tr M_1^l\right>
\eeq
it is a generating function for a disc of color $1$ (i.e. discrete surface with only one boundary of color $1$ and of length $l$), and $x$ is a complex "fugacity" conjugated to the boundary length $l$\footnote{Remark that these resolvents are properly defined when the fugacity $x \to \infty$}.

Like any expectation value in a formal matrix model \cite{BIPZ, ZJDFG}, it admits a topological $1/N^2$ expansion:
\beq
\ovl{W}_1(x) = \sum_{g=0}^\infty \ovl{W}_1^{(g)}(x)\,\, N^{1-2g}
\eeq
where $\ovl{W}_1^{(g)}(x)$ is the generating function for discrete surfaces of genus $g$.

The loop equations which allow to compute $\ovl{W}_1^{(g)}$ have been known for a long time \cite{staudacher}.
More recently, $\ovl{W}_1^{(g)}$ was computed for any $g$ in \cite{CEO, eyno, EOFg}.
The result for $\ovl{W}_1^{(0)}$ can be written in terms of an algebraic equation.
Let:
\beq
y(x) = V'_1(x) - \ovl{W}_1^{(0)}(x).
\eeq
$y(x)$ is solution of the following algebraic equation \cite{eyn1, eyn2}:
\beq
0=E(x,y(x))  = (V'_1(x)-y(x))(V'_2(y(x))-x) - P^{(0)}(x,y(x)) +1
\eeq
where
\beq
P(x,y) = \left<\Tr {V'_1(x)-V'_1(M_1)\over x-M_1}\,{V'_2(y)-V'_2(M_2)\over y-M_2}\right> = \sum_{g=0}^\infty N^{1-2g} P^{(g)}(x,y)
\eeq
and $y$ must be chosen as the branch of the solution of $E(x,y)=0$ which behaves like $V'_1(x)$ for large $x$.

\subsection{The spectral curve}

In general, correlation functions are multivalued functions of $x$, and it is better to write them as functions on a Riemann surface.

\smallskip

Therefore, we view $x$ and $y$ as two meromorphic functions living on a compact Riemann surface $\Sigma$.
\beq
E(x,y) = 0 \qquad \leftrightarrow \qquad \exists p\in\Sigma \,\, / \,\, x=x(p) \,\, {\rm and} \,\, y=y(p)
\eeq

Since the equation $E(x,y)=0$ has $\deg V_2$ solutions in $y$ for a given $x$, it means that for every point $p$ in $\Sigma$, there are $\deg V_2$ points $p^i$ in $\Sigma$ such that:
\beq\label{defsheetsx}
\forall i=0,\dots, d_2 \,\, , \qquad x(p^i) = x(p)
\eeq
where $d_2=\deg V'_2$, and by convention we assume $p^0=p$.

Similarly, if we regard $x$ as a function of $y$, then the equation $E(x,y)=0$ has $\deg V_1$ solutions for a given $y$, which means that for every point $p$ in $\Sigma$, there are $\deg V_1$ points $\td{p}^i$ in $\Sigma$ such that:
\beq\label{defsheetsy}
\forall i=0,\dots, d_1 \,\, , \qquad y(\td{p}^i) = y(p)
\eeq
where $d_1=\deg V'_1$, and by convention we assume $\td{p}^0=p$.

\subsection{Examples}

$\bullet$ If the algebraic curve $\Sigma$ build from $E(x,y)=0$ has genus zero, it is possible to find a rational parametrization \cite{eyn1, DKK}, i.e. $x(p)$ and $y(p)$ are rational functions of $p$:
\beq
\left\{
\begin{array}{l}
x(p) = {\gamma p} + \sum_{k=0}^{\deg V'_2} \alpha_k p^{-k} \cr
y(p) = {\gamma p^{-1}} + \sum_{k=0}^{\deg V'_1} \beta_k p^{k}
\end{array}
\right.
\eeq
where the coefficients $\alpha_k$, $\beta_k$ and $\gamma$ are determined by
$y(p) \sim_{p\to\infty} V'_1(x(p)) - 1/x(p) + O(p^{-2})$ and $x(p)\sim_{p\to 0} V'_2(y(p))-1/y(p) + O(p^2)$.

In that case the compact Riemann surface $\Sigma$ is the Riemann sphere.

\smallskip
This is the case which counts the Ising model bicolored maps.\\

$\bullet$ If the algebraic curve $\Sigma$ build from $E(x,y)=0$ has genus $1$, it is possible to find a parametrization with elliptical functions.

\smallskip

Spectral curves $E(x,y)=0$ of genus $g>0$, are not generating functions which counts maps, but they are still solutions of the loop equations, they have a more complicated combinatorical interpretation, and they are very useful for applications to string theory for instance.
In what follows, we assume that the spectral curve may have any genus, and one should keep in mind that only the genus zero case really corresponds to the Ising model on random surfaces.

%%%%%%%%%%%%%%%%%%%%%%%%%%%%%%%%%%%%%%%%%%%%%%%%%%%%%%%%%%%%%%%%%%%%%%%
\section{Definitions}

We assume that the spectral curve $E(x,y)=0$ is known, and that $x$ and $y$ are two meromorphic functions on the compact Riemann surface $\Sigma$.

\subsection{Notations}

The most general boundary condition for a discrete surface generated by the 2-matrix model is made of several boundaries, some of them having color $1$, some having color $2$, and some having mixed color boundaries.

Let us say that we have:
\begin{itemize}
\item $m$ boundaries of color $1$, with conjugated parameters $x(p_1),\dots,x(p_m)$,

\item $n$ boundaries of color $2$, with conjugated parameters $y(q_1),\dots,y(q_n)$,

\item $l$ mixed boundaries such that the $i^{\rm th}$ boundary is made of $2 k_i$ changes of colors.
It can be parameterized with $2 k_i$ conjugated length parameters $[x(p_{i,1}),y(q_{i,1}),x(p_{i,2}),y(q_{i,2}),x(p_{i,3}),y(q_{i,3}),\dots,x(p_{i,k}),y(q_{i,k})]$.

\end{itemize}
Notice that the $p_i$'s and $q_j$'s are points on the curve $\Sigma$.

\medskip

The generating function for discrete surfaces with that boundary condition is:
\bea\label{defH}
&& H_{k_1,\dots,k_l;m;n}(S_1,S_2, \dots, S_l;p_1 ,\dots, p_m;q_1,\dots,q_n)  \cr
&=&  \Big<
\prod_{i=1}^l  (N \delta_{k_i,1}+\Tr {1\over S_i})  \,\,
 \prod_{j=1}^m \Tr { 1 \over x(p_j)-M_1} \prod_{s=1}^n \Tr {1 \over y(q_s)-M_2}
\Big>_c \cr
&& + \delta_{l,0}\delta_{m,2}\delta_{n,0} {1\over (x(p_1)-x(p_2))^2}
 + \delta_{l,0}\delta_{m,0}\delta_{n,2} {1\over (y(q_1)-y(q_2))^2} \cr
&& + \delta_{l,0}\delta_{m,1}\delta_{n,0}  (y(p_1)-V'_1(x(p_1)))  + \delta_{l,0}\delta_{m,0}\delta_{n,1}  (x(q_1)-V'_2(y(q_1))) \cr
\eea
where
\beq
\Tr {1\over S_i} = \Tr \left({1 \over x(p_{i,1})-M_1}{1 \over y(q_{i,1})-M_2}
{1 \over x(p_{i,2})-M_1}
{1 \over y(q_{i,2})-M_2}
\dots
%{1 \over x(p_{i,k_i})-M_1}
{1 \over y(q_{i,k_i})-M_2}\right)
\eeq
and
\beq
S_i=[p_{i,1},q_{i,1},p_{i,2},q_{i,2},p_{i,3},q_{i,3},\dots,p_{i,k},q_{i,k}]
\eeq
is the ordered set of points $\{p_{i,l},q_{i,l}\}_{l=1 \dots k}$ up to cyclic permutations, i.e., using a graphical representation
\beq
S_i =
\begin{array}{r}
{\epsfxsize 4cm\epsffile{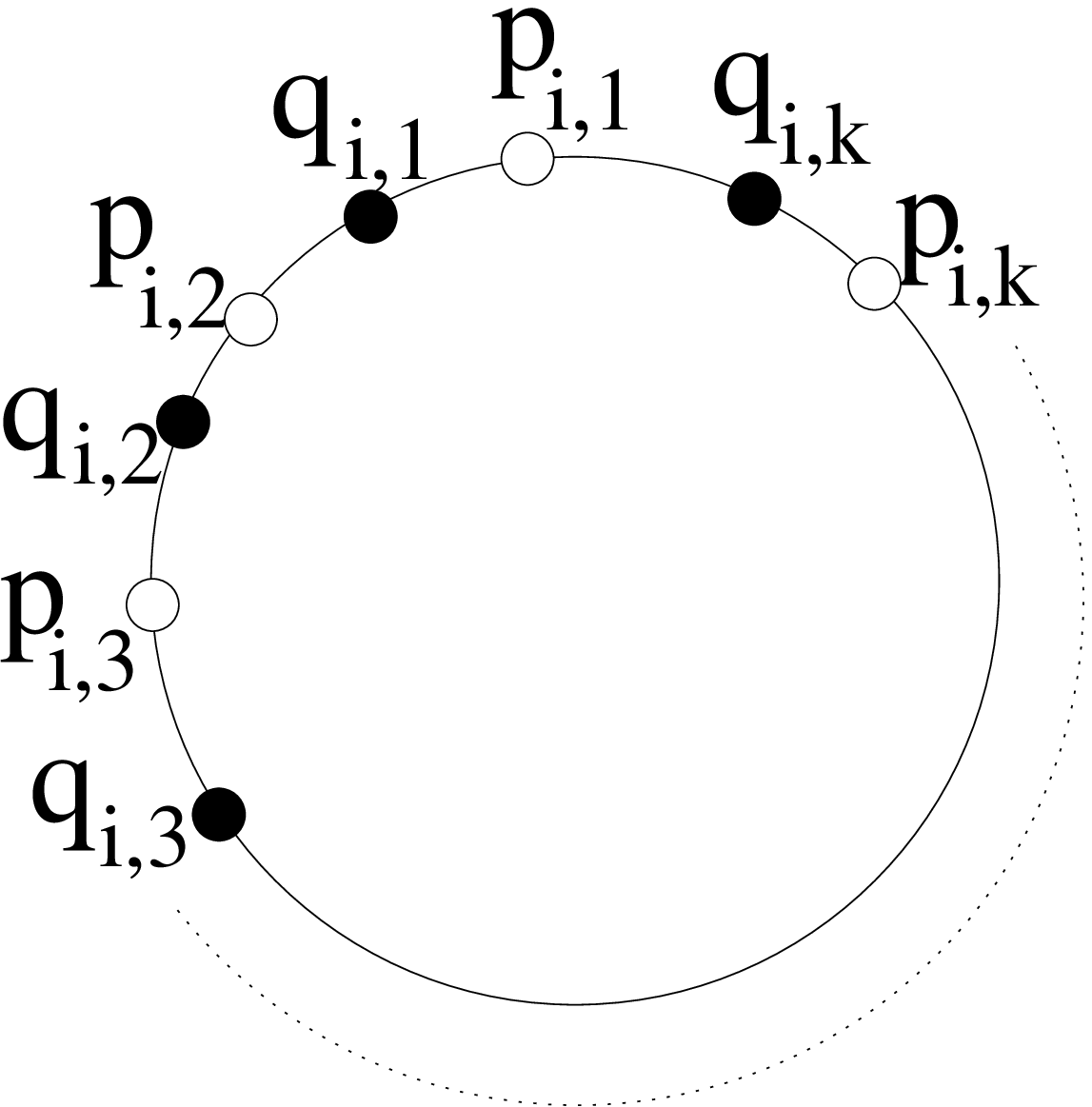}}
\end{array}.
\eeq
Each $p$ variable stands for a piece of boundary of color $1$, whereas each $q$ stands for a piece of color $2$.

\medskip

Each $H_{k_1,\dots,k_l;m;n}$ admits a topological expansion:
\beq
H_{k_1,\dots,k_l;m;n} = \sum_{g=0}^\infty N^{2-2g-l-m-n}\,\,H_{k_1,\dots,k_l;m;n}^{(g)}
\eeq
where $H_{k_1,\dots,k_l;m;n}^{(g)}$ is the generating function for discrete surfaces of genus $g$ with the same boundary conditions (indeed, the Euler characteristic of a surface of genus $g$ with $l+m+n$ boundaries is $\chi=2-2g-l-m-n$).

We represent $H_{k_1,\dots,k_l;m;n}^{(g)}$ graphically as a connected surface of genus $g$,
with $l$ circular boundaries, and $n+m$ punctures:
\beq
H_{{\bf k_L} ;m;n}^{(g)}({\bf S_L};p_1 ,\dots, p_m;q_1,\dots,q_n)
=
\begin{array}{r}
{\epsfxsize 6cm\epsffile{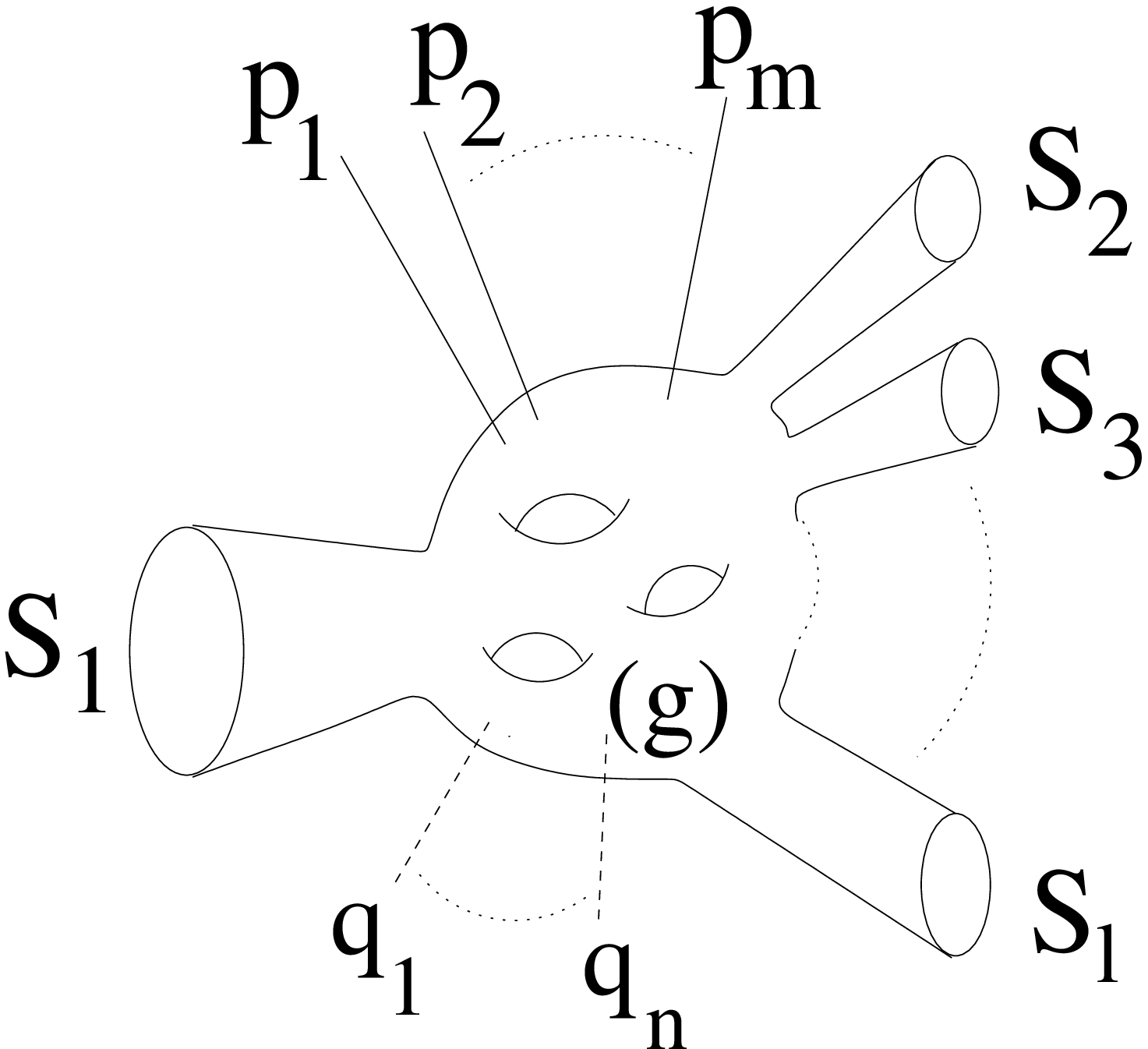}}
\end{array}.
\eeq

Since the correlation function $H_{k_1,\dots,k_l;m;n}^{(g)}$ does not depend on the order of the traces (i.e. one may permute the $S_i$'s),
we may choose one of the boundaries (for example $S_1$), and draw it on the exterior, and draw the whole surface in the interior of the circle $S_1$.
Moreover, because of the cyclic
invariance of the trace, we may choose a starting point on each boundary (for example $p_{1,1}$) by drawing an anticlockwise arrow on the boundary from this point\footnote{Remember that the
boundaries are oriented according to the sequence of points in the traces of the correlation functions.}.

Thus, we represent the correlation function $H_{{\bf k_L} ;m;n}^{(g)}({\bf S_L};p_1 ,\dots, p_m;q_1,\dots,q_n)$ by
a surface ${\cal S}_{{\bf k_L} ;m;n}^{(g)}({\bf S_L};p_1 ,\dots, p_m;q_1,\dots,q_n)$ which is a disc equipped with $g$ handles, $l-1$ holes corresponding to the $l-1$ remaining non homogenous boundaries,
$m$ white marked points corresponding to the homogenous boundaries of color $1$ and $n$ black marked
points corresponding to the homogenous boundaries of color $2$.
Note also that every non homogenous boundary is equipped
with a sequence of white and black points representing the sequence of boundary conditions.
\beq
H_{{\bf k_L} ;m;n}^{(g)}({\bf S_L};p_1 ,\dots, p_m;q_1,\dots,q_n)
=\begin{array}{r}
{\epsfxsize 6cm\epsffile{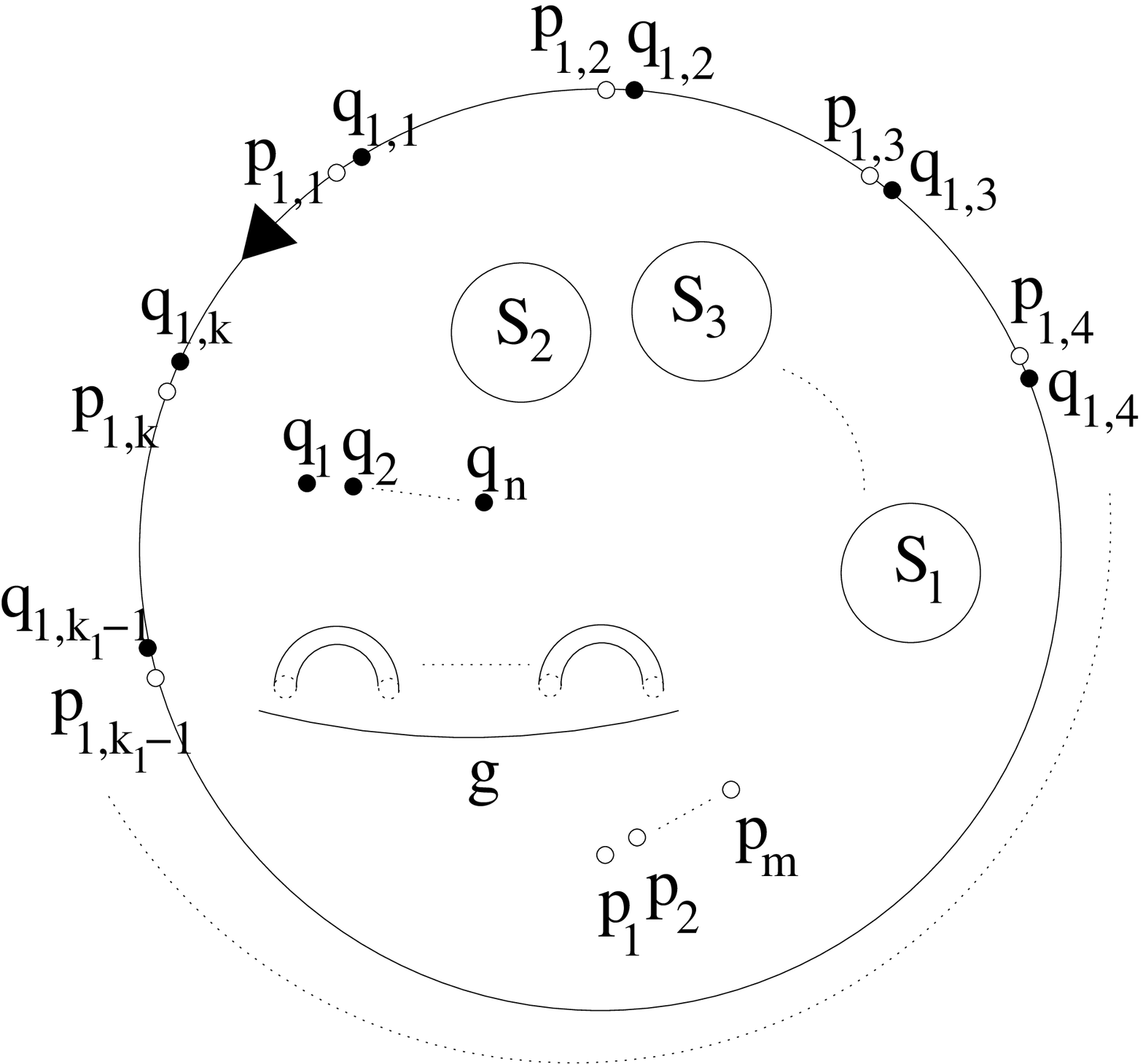}}.
\end{array}
\eeq
Notice that the other boundaries $S_2,\dots,S_l$ also have a marked edge $p_{i,1}\to q_{i,1}$ whose orientation is opposite (i.e. clockwise) of that of $S_1$.

\subsection{Previously known results}

Some cases are already known in the literature:

$\bullet$ {\bf Planar case:}
all $H_{k_1,\dots,k_l;0;0}^{(0)}$ (i.e. planar surfaces only) were computed in \cite{EObethe}.

$\bullet$ {\bf Non-mixed boundaries:}
all functions with only non-mixed boundaries, i.e. $H_{\emptyset;m;n}^{(g)}$ were computed in \cite{eyno, CEO, EOFg, EOsymFg}.

$\bullet$ {\bf Only one mixed boundary with $k=1$:}
$H_{1;m;n}^{(g)}$ was computed in \cite{EOsymFg}.

$\bullet$ {\bf In particular the sphere with one puncture is the resolvent:}
\beq
H^{(0)}_{0;1;0}(p) = \ovl{W}_1^{(0)}(p) = V'_1(x(p))-y(p)
\eeq

$\bullet$ {\bf In particular the sphere with one bicolored boundary is} \cite{DKK, eynmultimat, eyn1}:
\beq\label{H1000}
H^{(0)}_{1;0;0}(\{p,q\}) = {E(x(p),y(q))\over (x(p)-x(q))(y(q)-y(p))}
\eeq

\bigskip

Below, we compute all the other ones.

\section{Diagrammatic solution}

Here, we show the recipe to compute recursively any $H_{S_1,\dots,S_l;m;n}$.
The proof (which relies on loop equations, and is explained in the appendix is very technical, whereas the solution is rather simple and can be written pictorially.

\subsection{In equations}

In equations, the recursive solution of the loop equations (see the proof in appendix) can be written:
\beq\label{soluce}\encadremath{
\begin{array}{l}
H_{{\bf k_L} ;m;n}^{(g)}({\bf S_L};p_1 ,\dots, p_m;q_1,\dots,q_n) =\cr
{\displaystyle \Res_{r \to p_{1,1}, p_{i,\alpha},p_j, \tilde{q}_{1,k_1}^{j}} } { H_{1;0;0}^{(0)}(p_{1,1},q_{1,k_1})\, dx(r) \over
(x(p_{1,1})-x(r))(y(q_{1,k_1})-y(r))H_{1;0;0}^{(0)}(r,q_{1,k_1})} \,\,\times \cr
\Big\{
%%%%%%%%%%%%%%%%
%%%%%%%%%%%%%%%%
\sum_h \sum_{A\bigcup B=\{2, \dots,l\}}\sum_{\alpha =2}^{k_1} \sum_{I,J} H_{k_1-\alpha+1,{\bf k_B};m-\left|I\right|;n-\left|J\right|}^{(h)}(\{p_{1,\alpha},q_{1,\alpha},\dots p_{1,k_1},q_{1,k_1}\},{\bf S_B}; {\bf p_{M/I}};{\bf q_{N/J}})\cr
\qquad  \times { H_{\alpha-1,{\bf k_A};\left|I\right|;\left|J\right|}^{(g-h)}(\{r,q_{1,1},\dots p_{1,\alpha-1},q_{1,\alpha-1}\},{\bf S_A}; {\bf p_{I}};{\bf q_J})
 \over x(p_{1,\alpha})-x(r)} \cr
%%%%%%%%%%%%%%%%
%%%%%%%%%%%%%%%%
+\sum_{\alpha =2}^{k_1} { 1 \over x(p_{1,\alpha})-x(r)} \times \cr
 H_{\alpha-1, k_1-\alpha+1, { \bf k_{L/\{1\}}}; m;n}^{(g-1)}(\{r,q_{1,1},\dots p_{1,\alpha-1},q_{1,\alpha-1}\},\{p_{1,\alpha},q_{1,\alpha},\dots p_{1,k_1},q_{1,k_1}\}, {\bf S_{L/\{1\}}};{\bf p_M};{\bf q_N}) \cr
%%%%%%%%%%%%%%%%
%%%%%%%%%%%%%%%%
+ \sum_{i=2}^{l} \sum_{\alpha=1}^{k_i} {1 \over x(p_{i,\alpha}) - x(r)}  \times \cr
\quad \times  H_{k_1+k_i,{\bf k_{L/\{1,i\}}};m;n}^{(g)}( \{S_1(r),p_{i,\alpha},q_{i,\alpha},p_{i,\alpha+1}, \dots ,q_{i,k_i},p_{i,1},\dots ,p_{i,\alpha-1},q_{i,\alpha-1}\},{\bf S_{L/\{1,i\}}};{\bf p_M};{\bf q_N})  \cr
%%%%%%%%%%%%%%%%
%%%%%%%%%%%%%%%%
+ \sum_h \sum_{A\bigcup B=\{2, \dots,l\}}  \sum_{I,J} H_{k_1,{\bf k_A};\left|I\right|;\left|J\right|}^{(h)}(S_1(r),{\bf S_A}; {\bf p_{I}};{\bf q_J})
H_{{\bf k_B};m-\left|I\right|+1;n-\left|J\right|}^{(g-h)}({\bf S_B}; r, {\bf p_{M/I}};{\bf q_N/J})\cr
%%%%%%%%%%%%%%%%
%%%%%%%%%%%%%%%%
+\sum_{h=1}^{g} H_{0;1;0}^{(h)}(r) H_{k_1,\dots,k_l;m;n}^{(g-h)}(S_1(r),S_2, \dots, S_l;p_1 ,\dots, p_m;q_1,\dots,q_n) \cr

%%%%%%%%%%%%%%%%
%%%%%%%%%%%%%%%%
+ H_{{\bf k_L};m+1;n}^{(g-1)}({\bf S_K}(r);r,{\bf p_M};{\bf q_N})
\Big\} \cr
\end{array}}
\eeq
It looks terrible, but each term can be represented diagrammatically, and it is in fact rather simple and intuitive.
Let us notice for the moment that this formula involves residues (i.e. contour integrals on $\Sigma$) at various points, in particular the $\td{q}^j_{1,k_1}$ which are defined in eq.\ref{defsheetsy}, and were we mean $j\neq 0$.

This formula also involves the function $H_{1;0;0}^{(0)}$ which is given in eq.\ref{H1000}.

All the other terms in the RHS of eq.\ref{soluce} are either some $H_{S;m;n}^{(g)}$'s computed recursively by the same formula, or some $H_{0;m;n}^{(g)}$ which were computed in \cite{eyno, CEO, EOFg, EOsymFg}.

\subsection{Diagrammatic representation}

It is more convenient to represent equation \ref{soluce} diagrammatically:
\bea
\begin{array}{r}
{\epsfxsize 4.5cm\epsffile{mix1.eps}}
\end{array}
= \begin{array}{r}
{\epsfxsize 4.5cm\epsffile{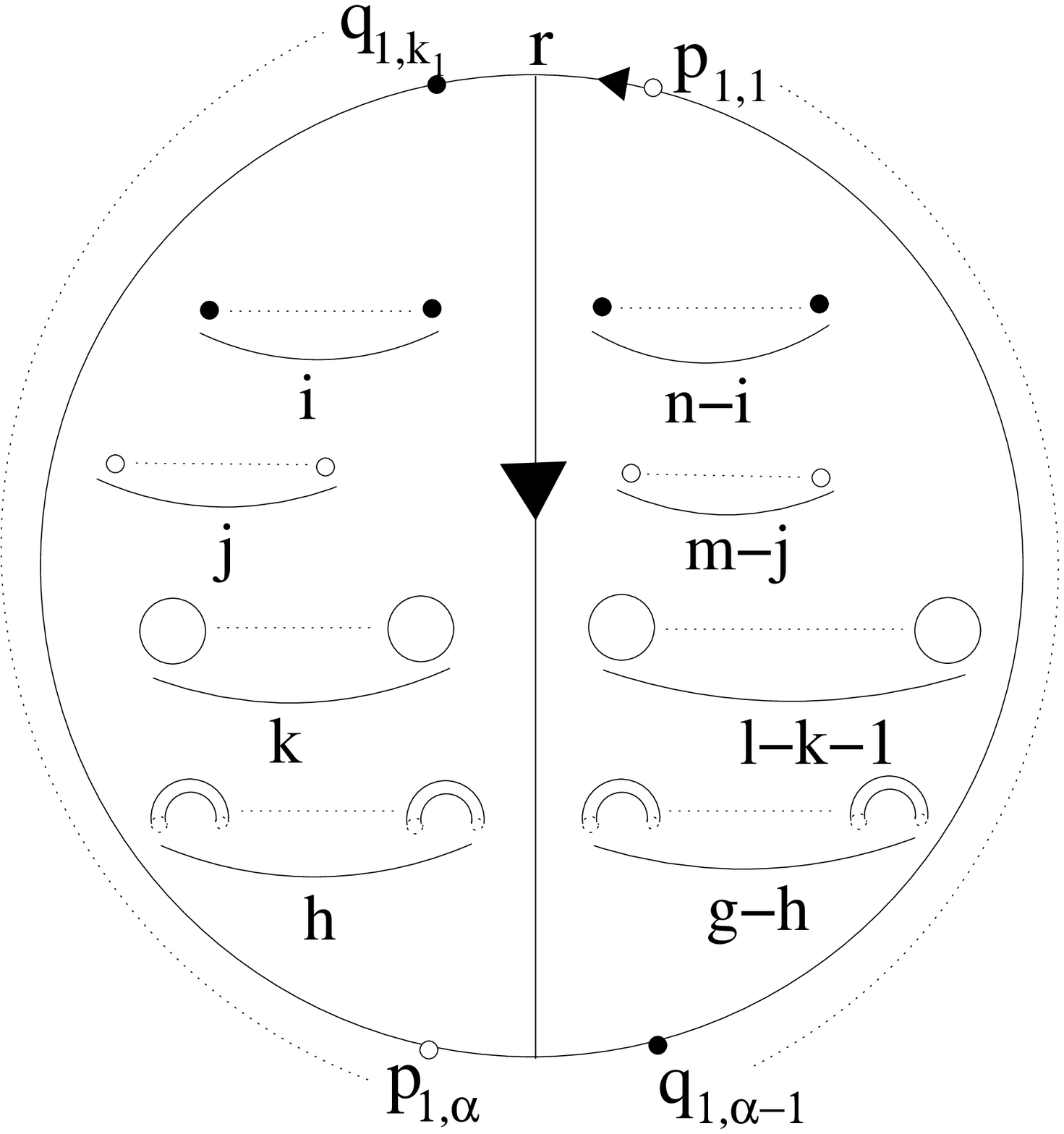}}
\end{array}
+\begin{array}{r}
{\epsfxsize 4.5cm\epsffile{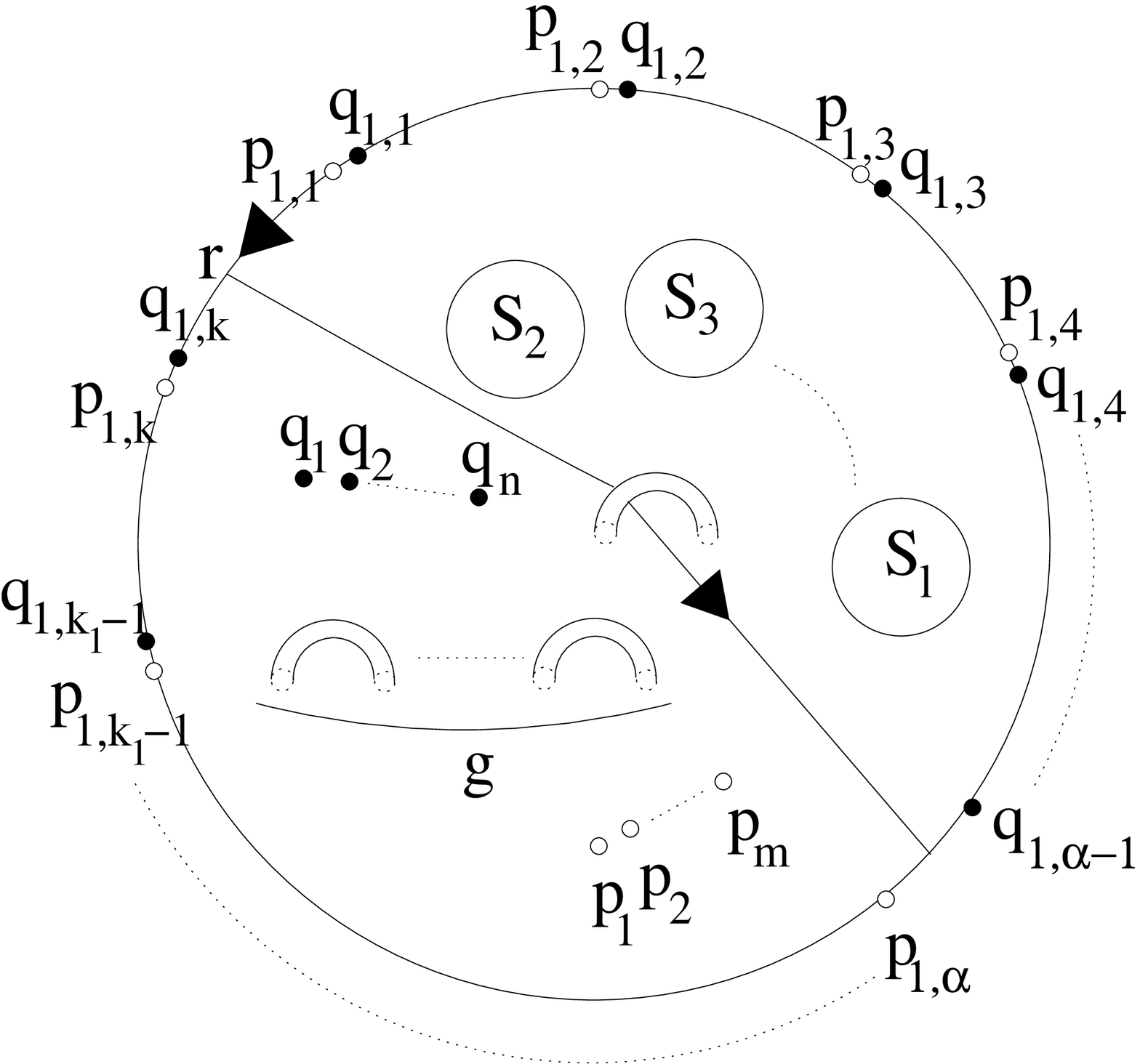}}
\end{array}
\cr
+
\begin{array}{r}
{\epsfxsize 4.5cm\epsffile{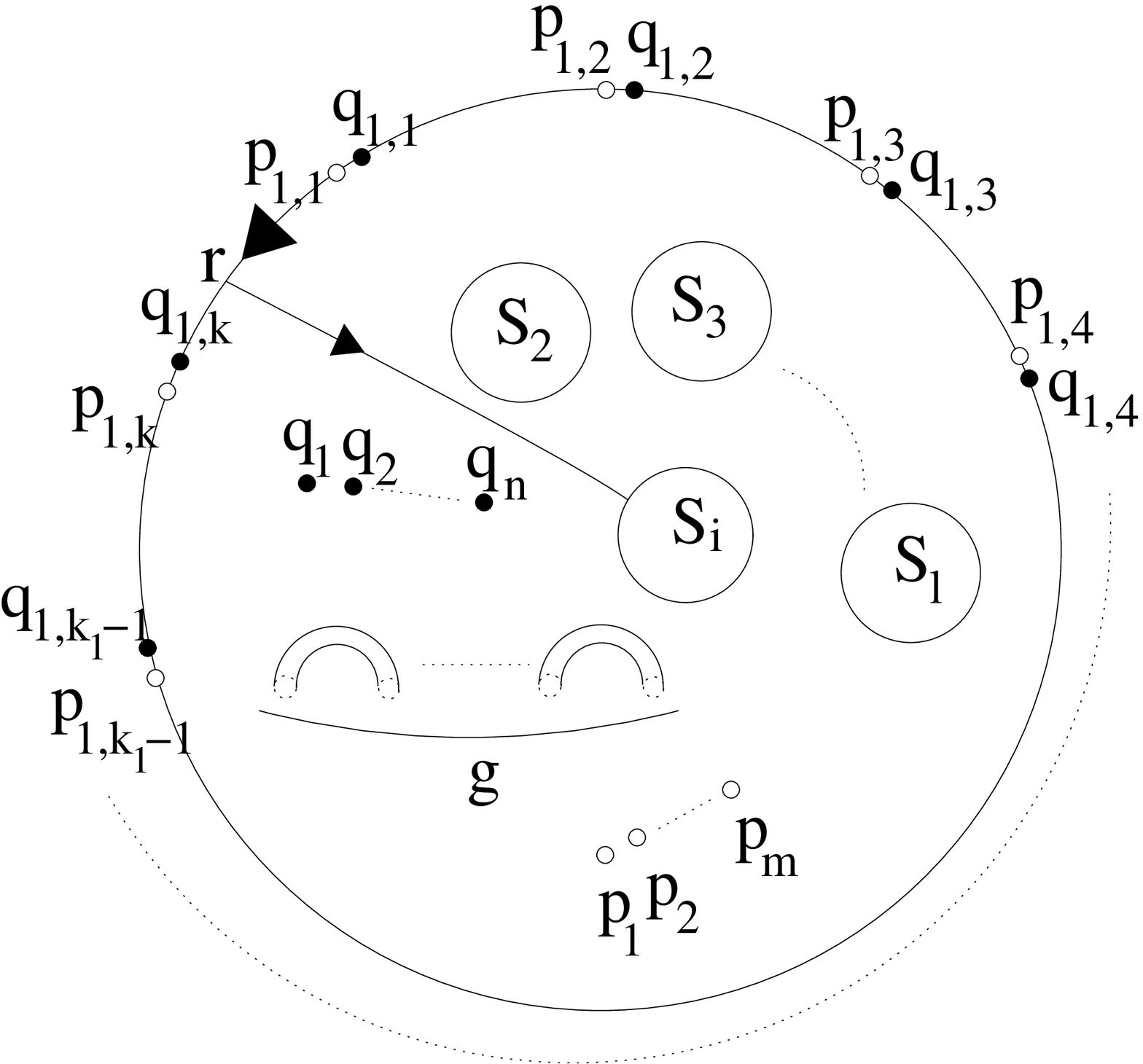}}
\end{array}
+\begin{array}{r}
{\epsfxsize 4.5cm\epsffile{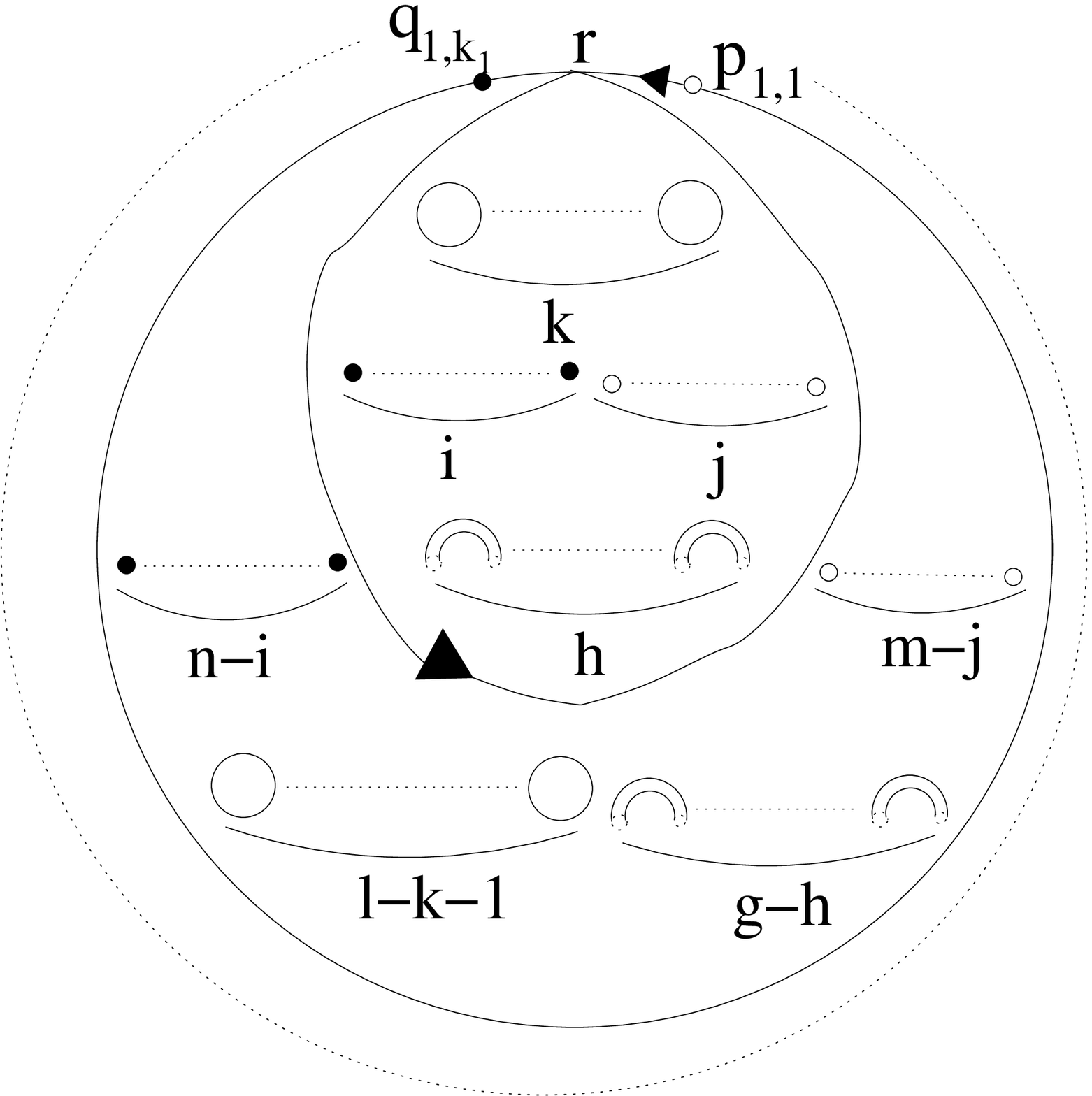}}
\end{array}
+\begin{array}{r}
{\epsfxsize 4.5cm\epsffile{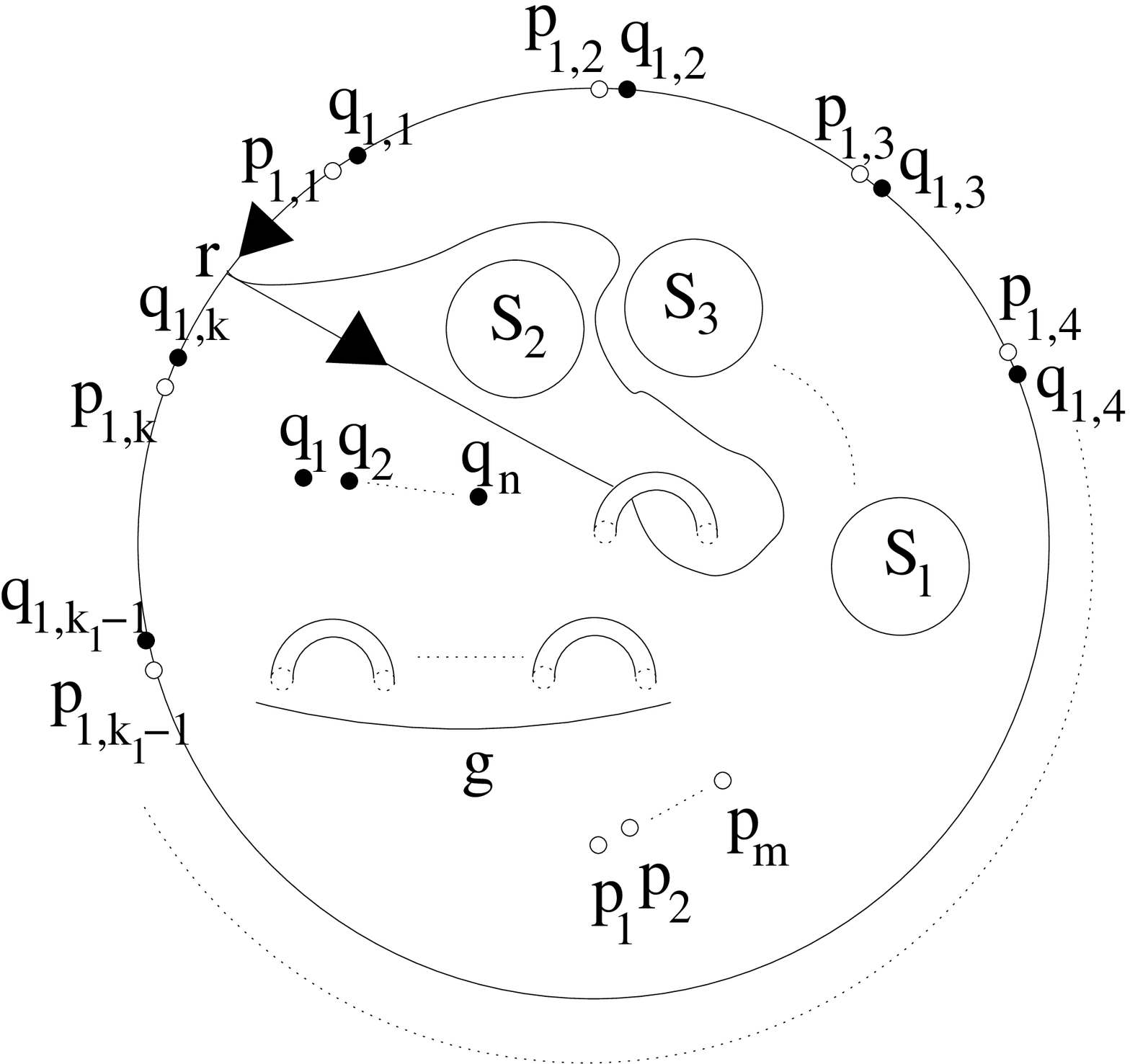}}
\end{array}.
\cr
\eea
where we explain the meaning of those graphs below.

\subsubsection{Cutting surfaces}

Consider a connected surface ${\cal S}$ with at least one boundary (i.e. $l\geq 1$):
\beq
{\cal S} =
{\cal S}_{{\bf k_L} ;m;n}^{(g)}({\bf S_L};p_1 ,\dots, p_m;q_1,\dots,q_n)
=\begin{array}{r}
{\epsfxsize 6cm\epsffile{mix1.eps}}.
\end{array}
\eeq

Let ${\rm Cut}({\cal S})$ be the set of all topologically inequivalent possibilities of cutting the surface along a line $p_{1,1}\to p_{i,\alpha}$ (we allow the closed line $(i,\alpha)=(1,1)$).
When we cut along such a line, we can either get a connected or a disconnected surface.
The only possibility of getting a disconnected surface is if the point $p_{i,\alpha}$ belongs to $S_1$, i.e. $i=1$, and if there is no handle going above the cut.

\medskip

Here is the algorithm to construct ${\rm Cut}({\cal S})$:

\smallskip

\begin{itemize}
\item
%
%Let us now extend this set of graphs by dressing the surface with a path starting from the starting point and going to another
%boundary.
%For this purpose, consider a bare surface ${\cal S}_{{\bf k_L} ;m;n}^{(g)}({\bf S_L};p_1 ,\dots, p_m;q_1,\dots,q_n)$
%(remember that it means that one has specified one outer boundary $S_1$ and one starting point $p_{1,1}$).
%To dress it,
one first has to choose any ending point $p_{i,\alpha}$ on a mixed
boundary and draw a path going from the left of the starting point to the left of the ending point\footnote{The orientation is seen from the
point of view of an observer living on the upper side of the disc.}:
\begin{itemize}

\item This point can belong to a boundary different from the starting one, $i \neq1$:
\beq
\begin{array}{r}
{\epsfxsize 4.5cm\epsffile{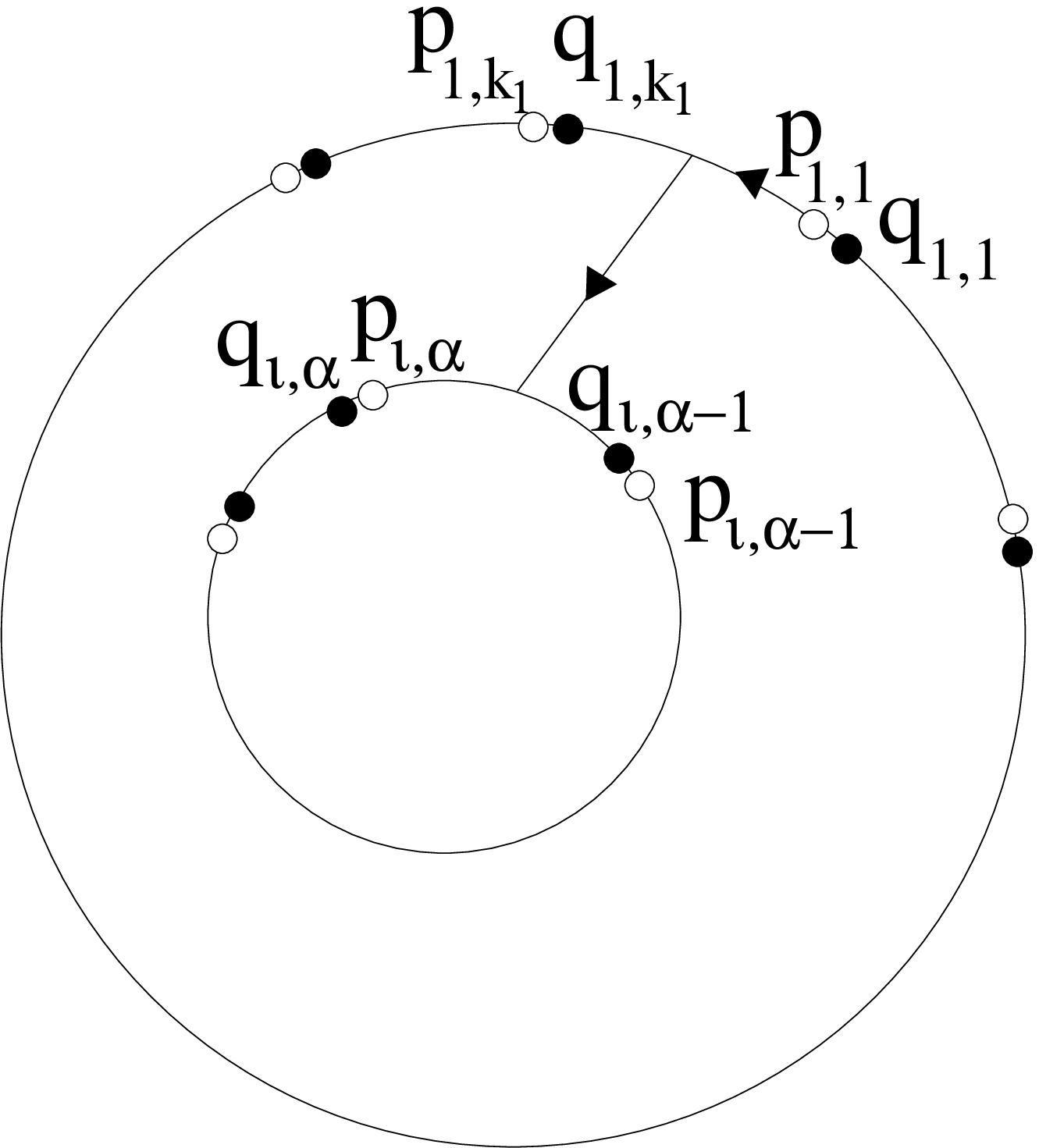}}.
\end{array}
\eeq
There are $\sum_{i =2}^l k_i$ such possibilities.

\item It can belong to the same boundary, $i = 1, \alpha\neq 1$ :
\beq
\begin{array}{r}
{\epsfxsize 4.5cm\epsffile{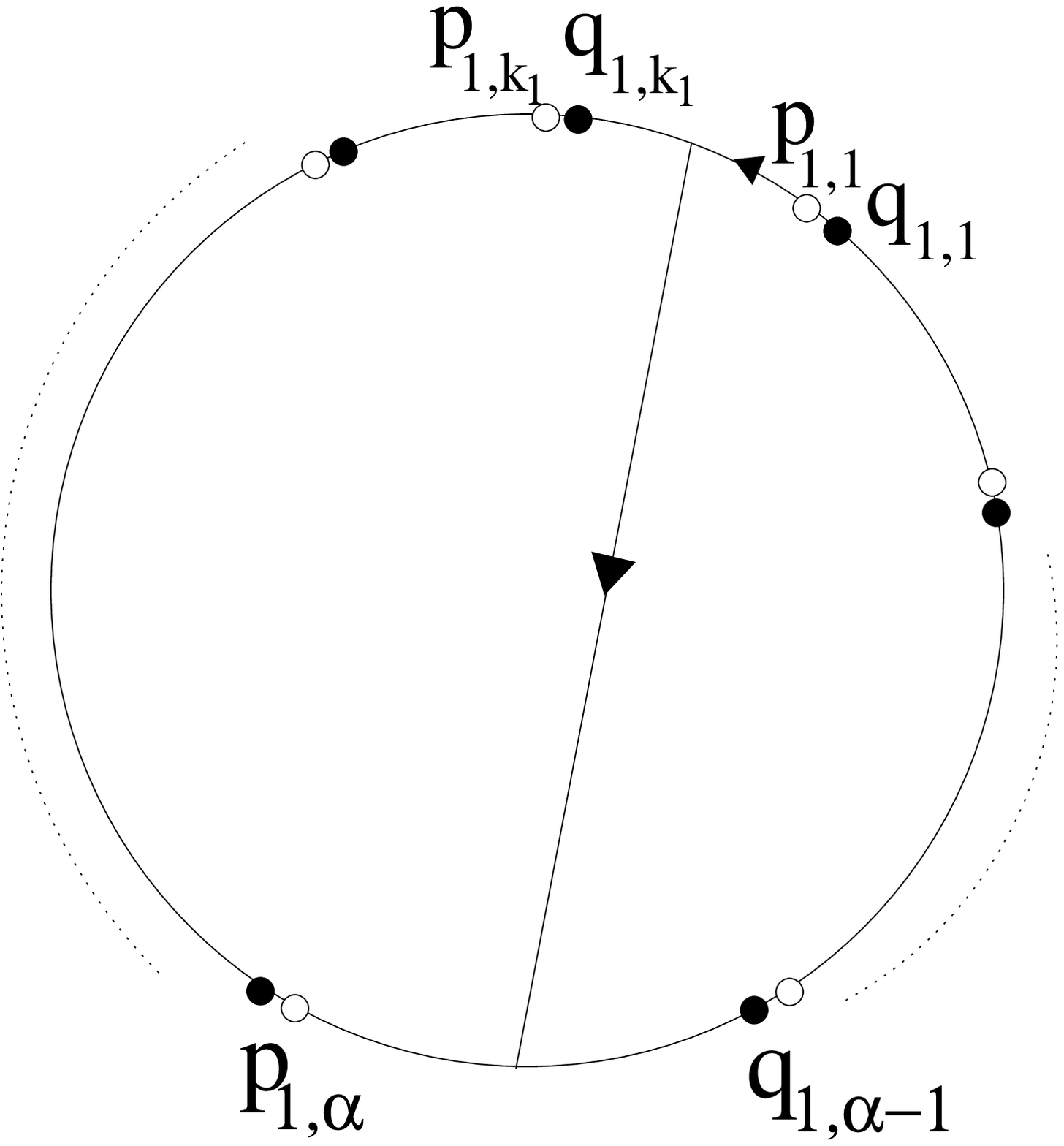}}
\end{array}
\eeq
There are $k_1-1$ such possibilities.

\item It can be the same as the starting point $i=1,\alpha=1$:
\beq
\begin{array}{r}
{\epsfxsize 4.5cm\epsffile{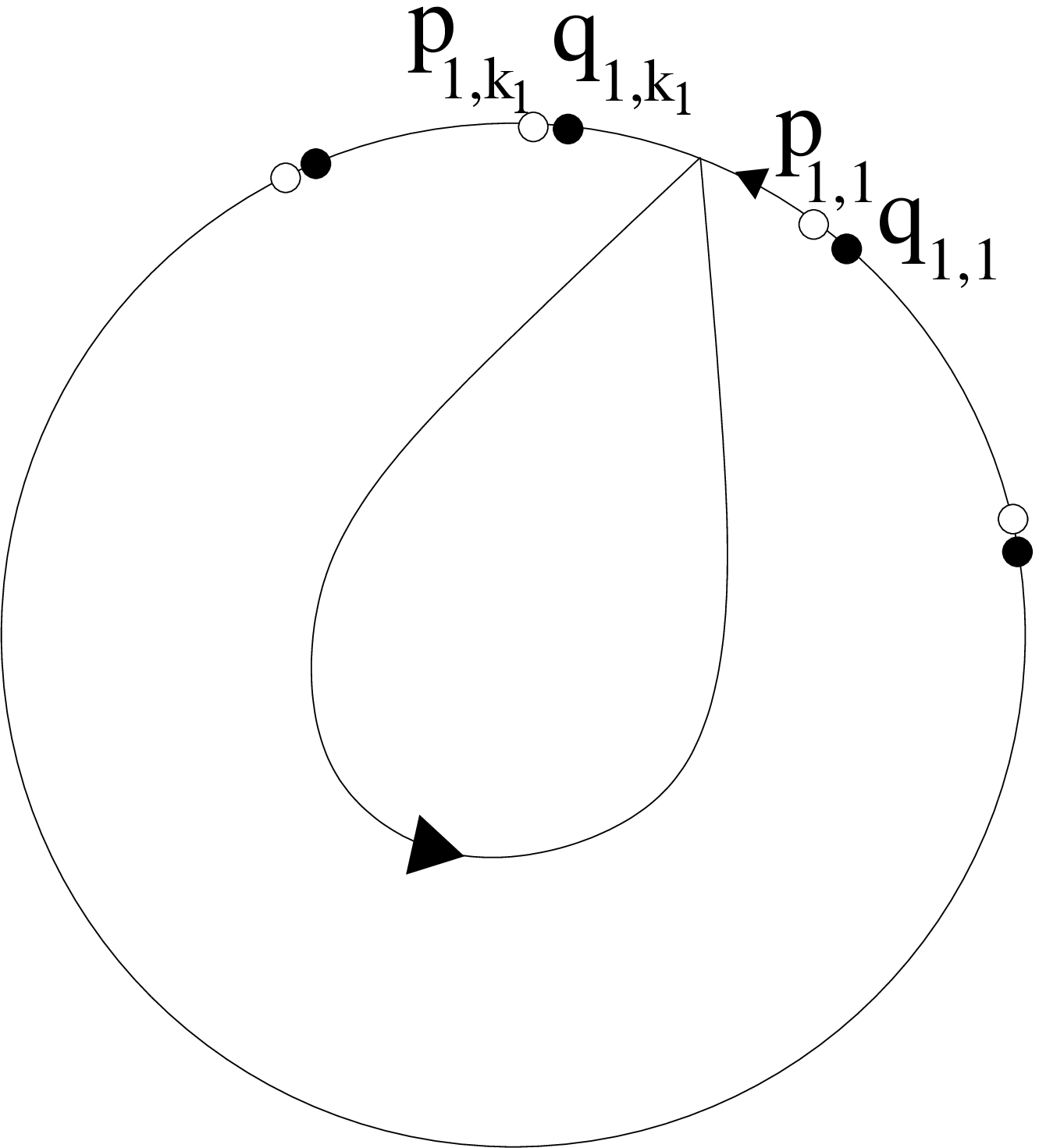}}
\end{array}
\eeq
There is only one such possibility.

\end{itemize}

\item Once this ending point is chosen, it remains to fix the position of the handles and the other boundaries and punctures with respect to this
path. The number of inequivalent possibilities depends on the respective position of the starting and ending points:
\begin{itemize}

\item If the starting and ending points do not belong to the same boundary, the surface is not disconnected by the cut, and every choices are equivalent since the left and right
side of the path belong to the same component of the surface:
\beq
\begin{array}{r}
{\epsfxsize 4.5cm\epsffile{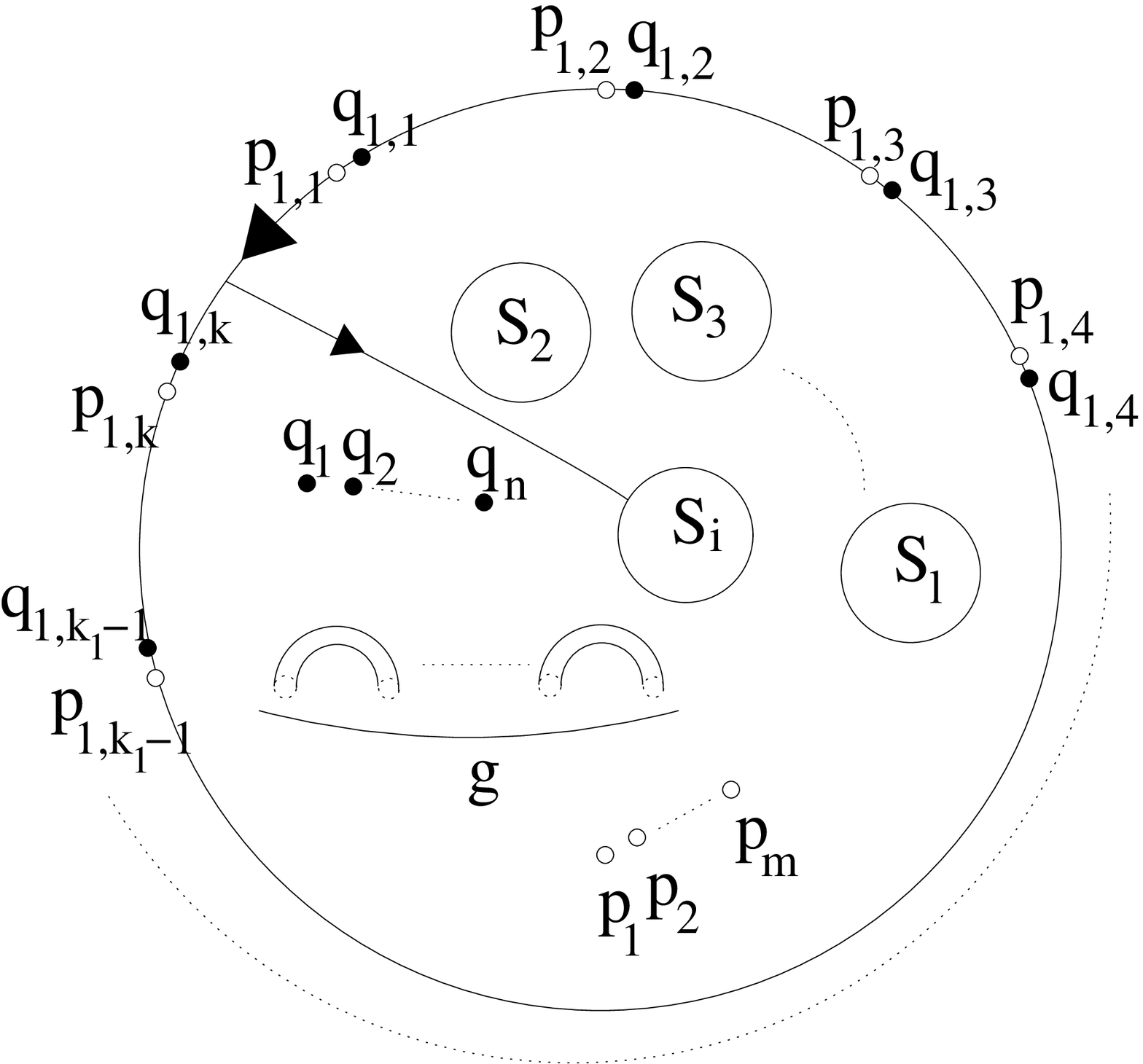}}
\end{array}
\eeq
There is only one possibility for the boundaries, punctures and handles configuration.

\item If the starting and ending points belong to the same boundary, two different configurations can occur: either the path
does cut the disc into two disconnected parts, i.e. no handle goes above the path. In this case, one has to choose for each handle
and boundary whether it lies to the left or the right of the path:
\beq
\begin{array}{r}
{\epsfxsize 4.5cm\epsffile{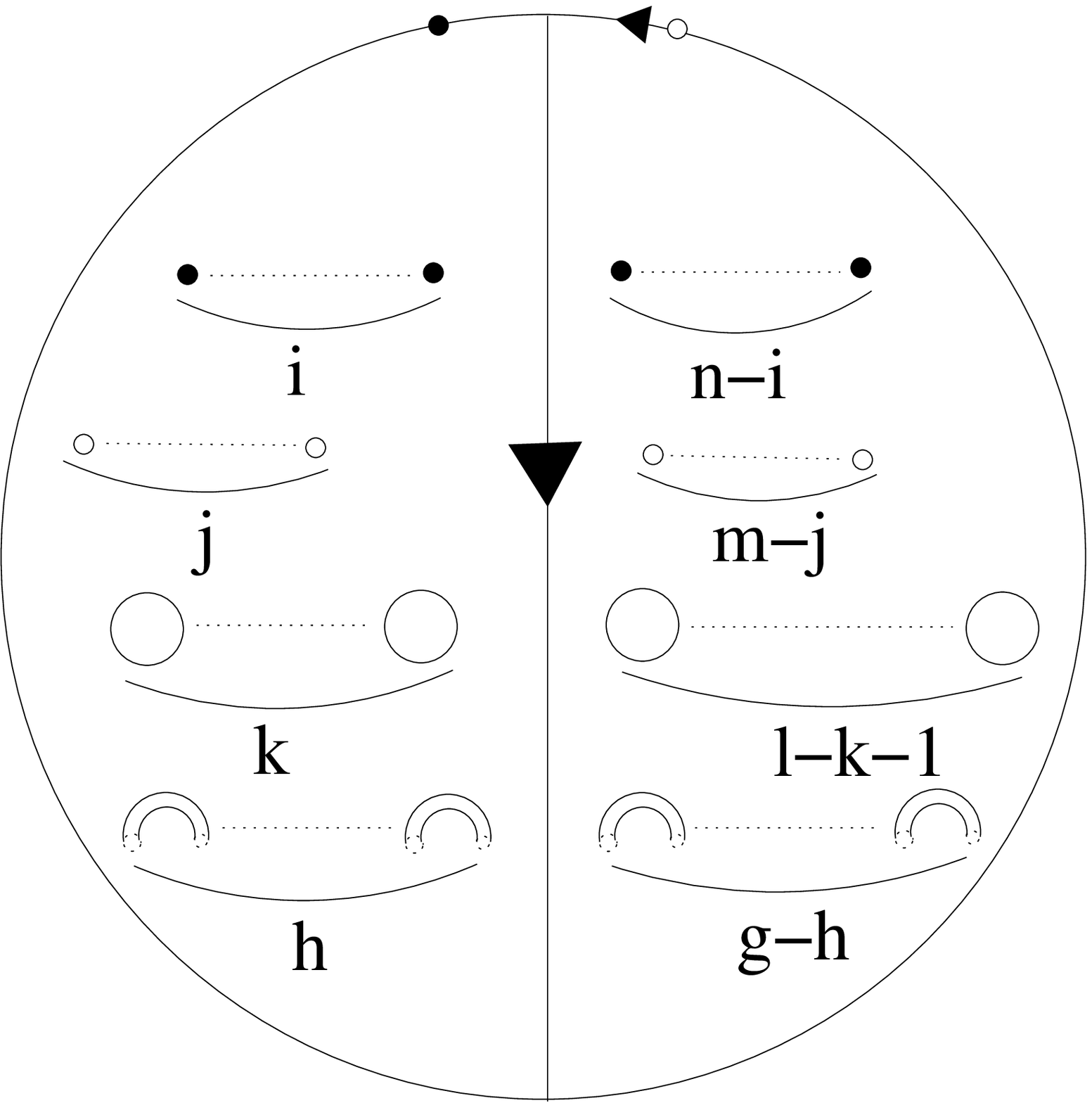}}
\end{array}
\eeq
There are $2^{n+m+g}$ such configurations;

Either the path does not separate the disc into two parts, and all the positions of handles, punctures and boundaries are equivalent:
\beq
\begin{array}{r}
{\epsfxsize 4.5cm\epsffile{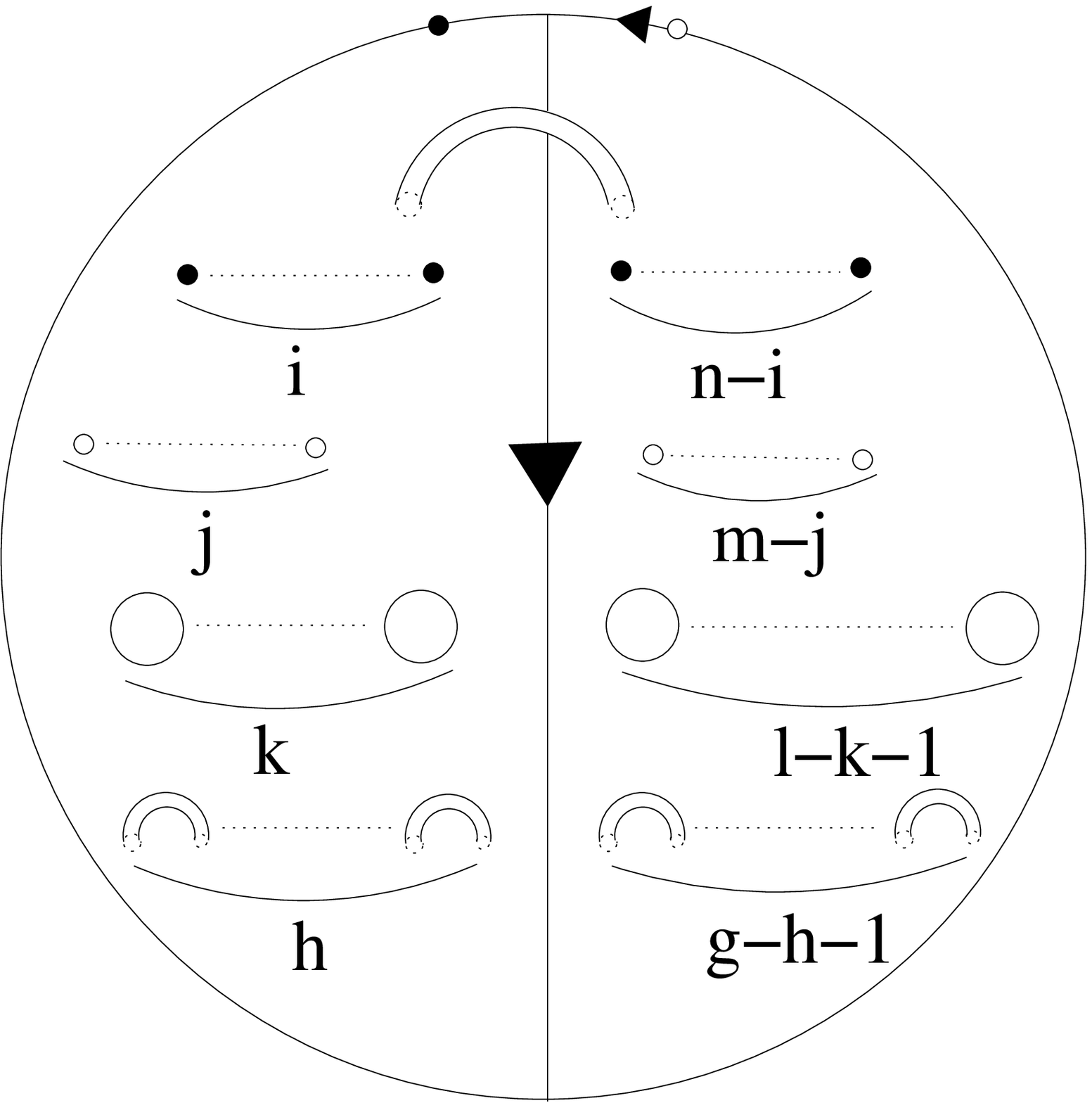}}
\end{array}
\eeq
There is only one such configuration because one can transport the handles, punctures and boundaries across the handle above
the path.

\end{itemize}

\end{itemize}

We have then built the set ${\rm Cut}({\cal S})$ of cut surfaces associated to any surface ${\cal S}$.

%\subsubsection{Equivalence class of bare surfaces}

%\bd
%Two graphs ${\cal S}_{{\bf k_L} ;m;n}^{(g)}({\bf S_L};p_1 ,\dots, p_m;q_1,\dots,q_n)$
%and ${\cal \tilde{S}}_{{\bf k_L} ;m;n}^{(g)}({\bf \tilde{S}_L};\tilde{p}_1 ,\dots, \tilde{p}_m;\tilde{q}_1,\dots,\tilde{q}_n)$ are equivalent if and only if:
%\begin{itemize}
%\item The homogenous boundaries are the same: $\{p_i\}_{i=1}^m = \{\tilde{p}_i\}_{i=1}^m$ and $\{q_i\}_{i=1}^n = \{\tilde{q}_i\}_{i=1}^n$

%\item The non-homogenous boundaries are the same up to cyclic permutation of their elements: $\{S_i\}_{i=1}^l \cong \{\tilde{S}_i\}_{i=1}^l$
%where $S_i \cong S_j$ if they coincide up to cyclic permutation of their elements.
%\end{itemize}

%We note $\left[ {\cal S} \right]$ the equivalent class of a graph ${\cal S}$.

%\ed

%This means that two graphs are equivalent if they represent the same surface but with different outer boundary and starting point.
%The equivalent class of a graph ${\cal S}$ can thus be viewed as surface described by ${\cal S}$ but without any particular view
%point:
%\beq
%\left[
%\begin{array}{r}
%{\epsfxsize 3cm\epsffile{mix1.eps}}
%\end{array}
%\right]
%=
%\begin{array}{r}
%{\epsfxsize 3cm\epsffile{blobmix1.eps}}.
%\end{array}
%\eeq

\subsection{Weights of graphs}

Now, let us associate a weight to each cut surface.
We define recursively a weight ${\cal P}$ on the set of graphs:
\bd
The weight ${\cal P}$ of an uncut surface is given by the corresponding correlation function:
\beq
{\cal P}\left( {\cal S}_{{\bf k_L} ;m;n}^{(g)}({\bf S_L};p_1 ,\dots, p_m;q_1,\dots,q_n)\right)
:= H_{{\bf k_L} ;m;n}^{(g)}({\bf S_L};p_1 ,\dots, p_m;q_1,\dots,q_n).
\eeq

%\beq
%{\cal P}\left( {\cal S}_{m;{\bf k_L} ;n}^{(g)}(p_1 ,\dots, p_m;{\bf S_L};q_1,\dots,q_n)\right)
%:= H_{{\bf k_L} ;m;n}^{(g)}({\bf S_L};p_1 ,\dots, p_m;q_1,\dots,q_n).
%\eeq

The weight of the disconnected union of two surfaces is the product of their respective weights:
\beq
{\cal P}({\cal S} \bigcup {\cal S'} ):= {\cal P}({\cal S}) \times {\cal P} ({\cal S'} ).
\eeq

The weight of a cut surface is obtained from the weight of the surface(s) obtained by cutting along the
path $\gamma$ following the rules:
\begin{itemize}
\item If the starting and ending points of $\gamma$ do not coincide:
\beq\label{recd2}
\CP\left(
\begin{array}{r}
{\epsfxsize 1.5cm\epsffile{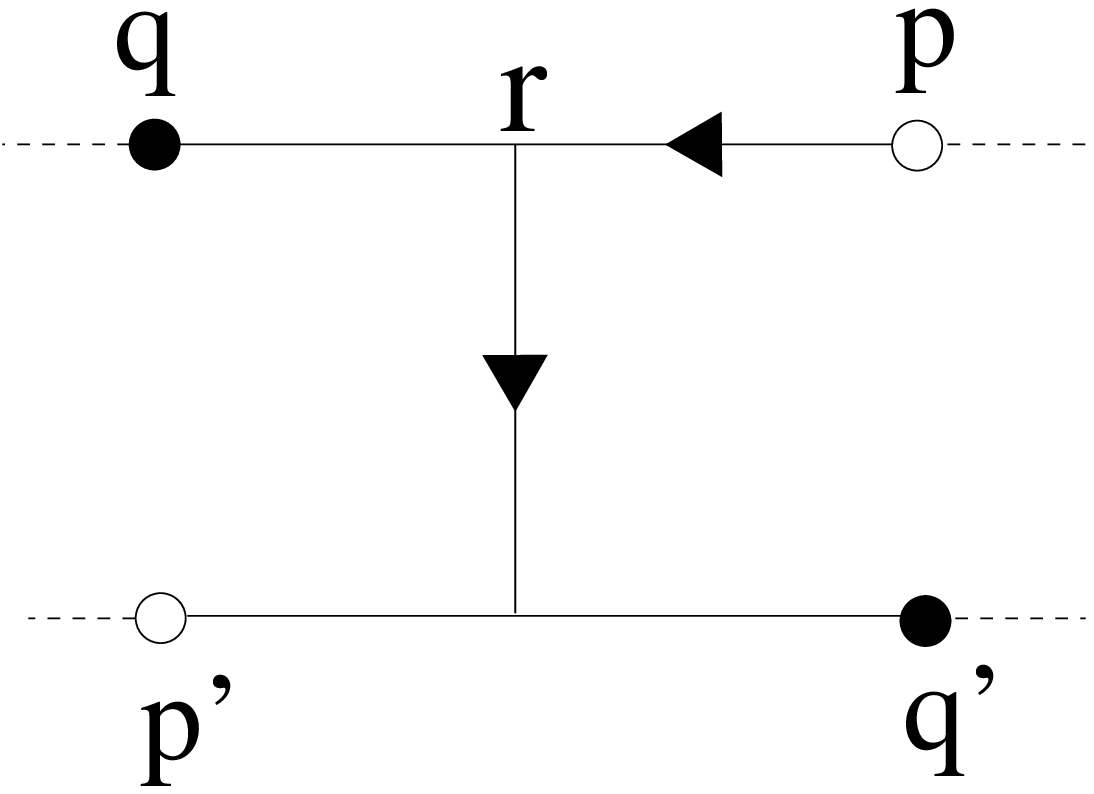}}
\end{array}\right) =
\Res_{r \to p, \tilde{q}^j,p'}
{ H_{1;0;0}^{(0)}(p,q) \over
(x(p)-x(r))(y(q)-y(r))(x(p')-x(r)) H_{1;0;0}^{(0)}(r,q)}
\CP \left(\begin{array}{r}
{\epsfxsize 1.5cm\epsffile{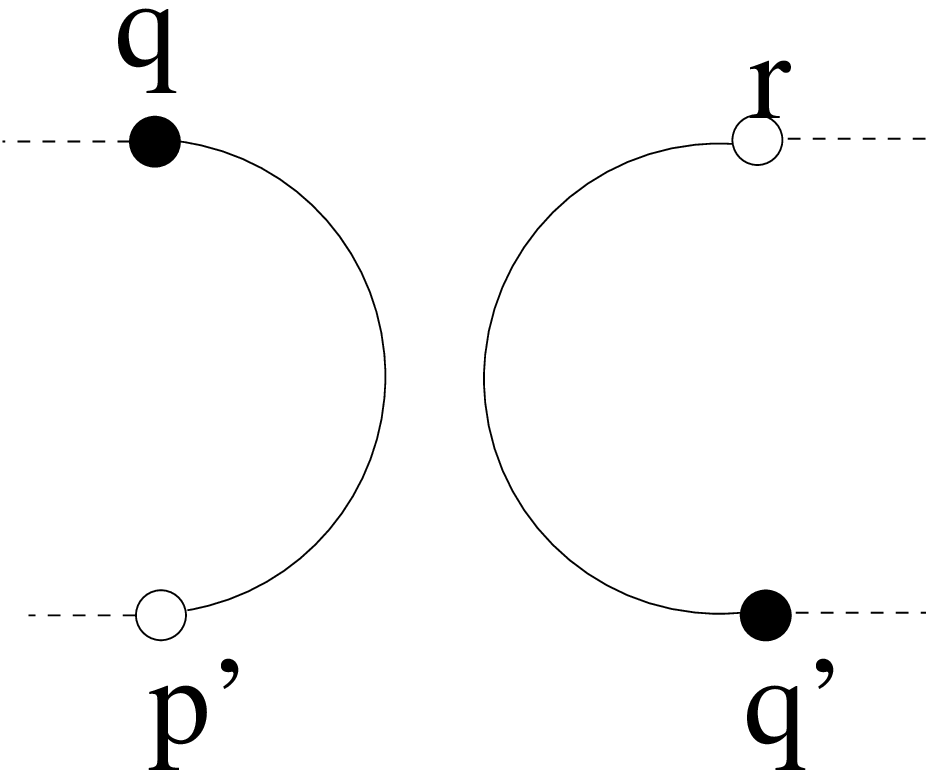}}
\end{array} \right)
\eeq

\item If the starting and ending points coincide:
\beq\label{recd3}
\CP\left(
\begin{array}{r}
{\epsfxsize 2cm\epsffile{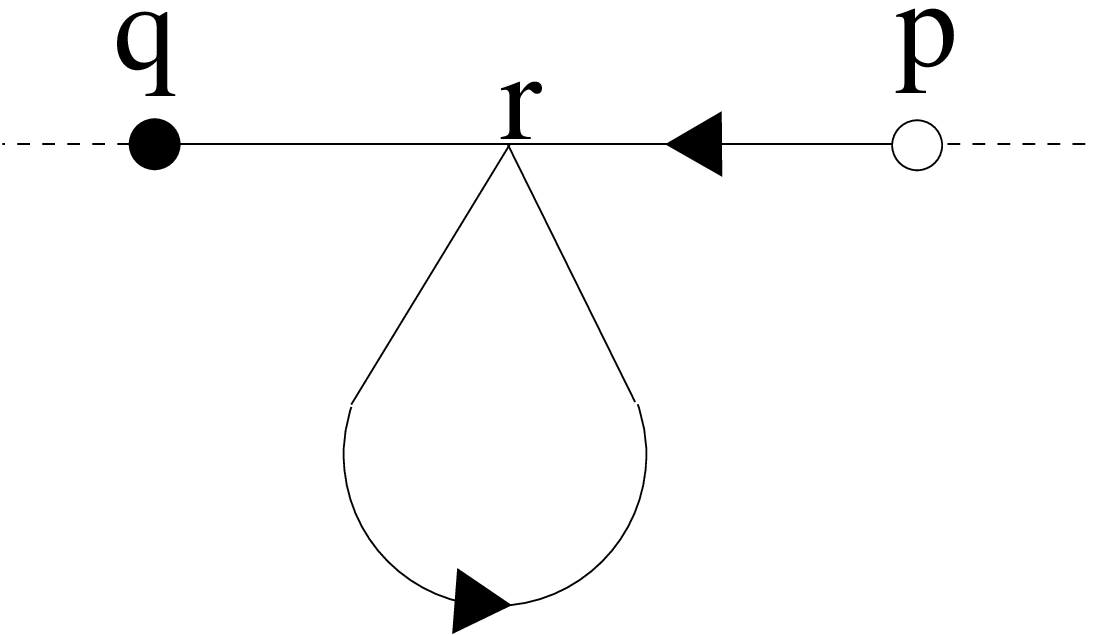}}
\end{array}\right) =
 \Res_{r \to p, \tilde{q}^j, p_i}
{ H_{1;0;0}^{(0)}(p,q) \over
(x(p)-x(r))(y(q)-y(r)) H_{1;0;0}^{(0)}(r,q)}
\CP \left(\begin{array}{r}
{\epsfxsize 2cm\epsffile{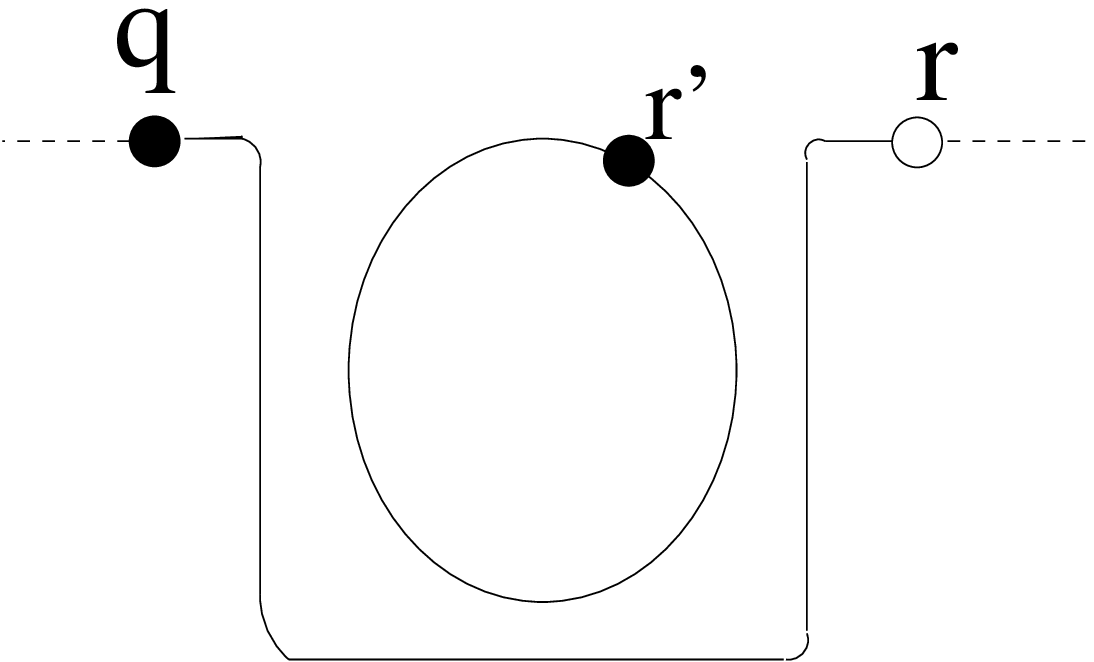}}
\end{array} \right)
\eeq
where the $p_i$'s are the points encircled inside the closed loop.
\end{itemize}

\ed

%\br
%Thanks to the symmetry of the correlation functions, two equivalent bare surfaces have the same weight, i.e. we have
%defined a measure on the set of equivalence class defined in the preceding paragraph.
%\er

With such notations, equation \ref{soluce} can be reinterpreted as:
\bt
The weight of a given surface is equal to the sum of the weights of all corresponding cut surfaces:
\beq
{\cal P}({\cal S}) = \sum_{S\in {\rm cut}({\cal S})} {\cal P} (S)
\eeq
\et
%
%\proof{
%This is just the recursion relation \eq{soluce}.
%}
%
I.e. graphically:
\bea
\begin{array}{r}
{\epsfxsize 4.5cm\epsffile{mix1.eps}}
\end{array}
= \begin{array}{r}
{\epsfxsize 4.5cm\epsffile{soluce2.eps}}
\end{array}
+\begin{array}{r}
{\epsfxsize 4.5cm\epsffile{soluce4.eps}}
\end{array}
\cr
+
\begin{array}{r}
{\epsfxsize 4.5cm\epsffile{soluce3.eps}}
\end{array}
+\begin{array}{r}
{\epsfxsize 4.5cm\epsffile{soluce1.eps}}
\end{array}
+\begin{array}{r}
{\epsfxsize 4.5cm\epsffile{soluce5.eps}}
\end{array}.
\cr
\eea

Performing this procedure recursively on any correlation functions, one can eliminate the mixed boundaries step by step
until there is no mixed boundary left, i.e. until there are only punctures left.
The correlation functions with only punctures are computed in \cite{eyno, CEO, EOFg}.

\section{Examples of applications}

In this section, we show how to use our formula to recover some previously known results, in particular the planar case, and surfaces with uniform boundaries.
We also compute two simple examples: the generating function of discs with four boundary
operators and the generating function of cylinders with two boundary operators on each boundary.

\subsection{Link with former results}

\subsubsection{Planar mixed traces}

If one is interested in the planar mixed correlation functions with only one boundary, the recursion relation simplifies to:
\beq
\begin{array}{r}
{\epsfxsize 4.5cm\epsffile{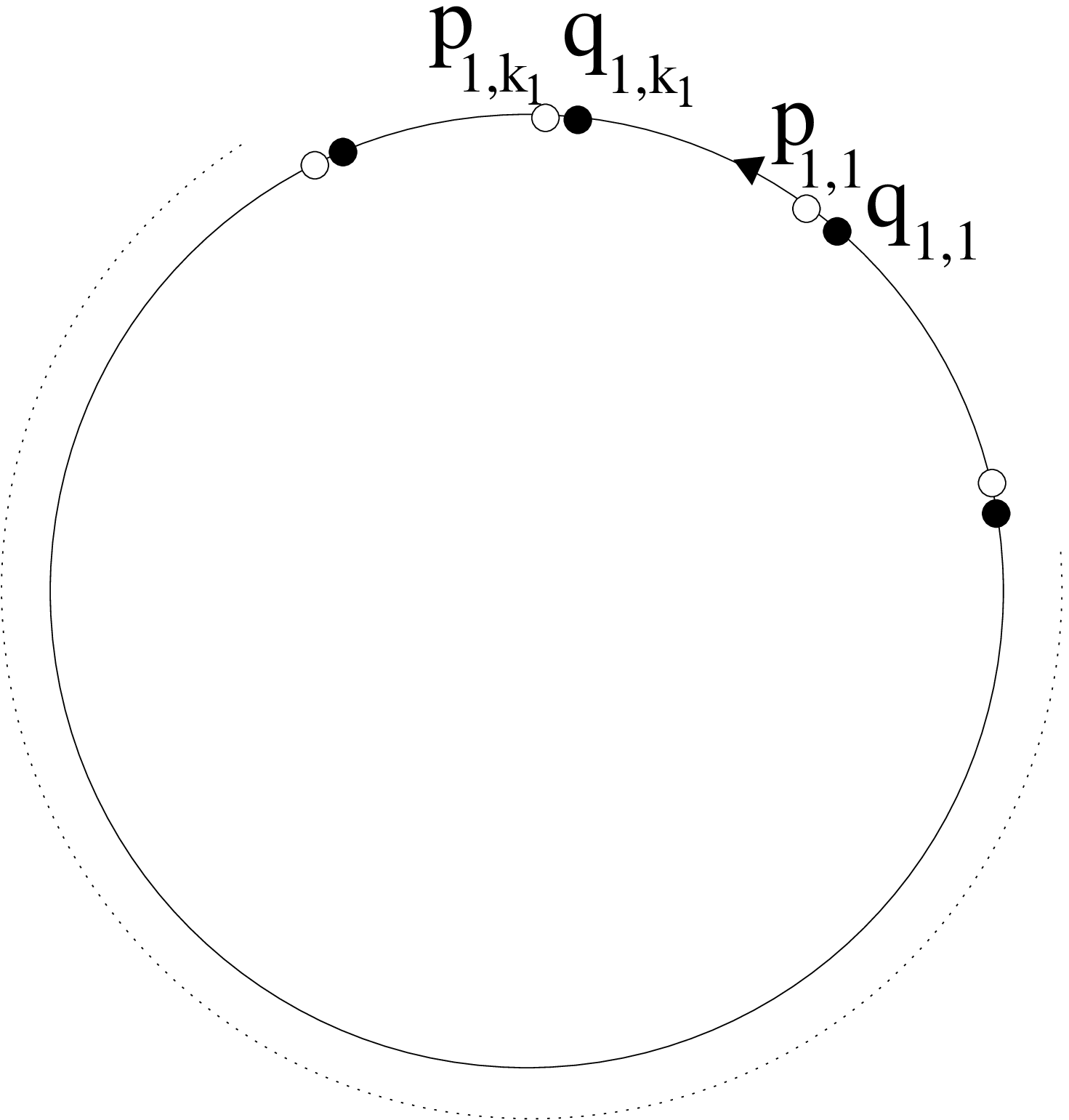}}
\end{array}
 =
 \sum_{\alpha = 2}^{k_1}
 \begin{array}{r}
{\epsfxsize 4.5cm\epsffile{path2.eps}}
\end{array}
\eeq

One can thus draw the result of the whole recursive procedure as the sum over all possible link patterns on the starting disc
in such a way they separate all boundary variables. This reproduce the decomposition used in \cite{EObethe} to compute
the building blocks $F_k = C_{id}^{k}$.

{\bf Example:}

The three point mixed correlation function reads:

\beq
H_{3;0;0}^{(0)}(p_1,q_1,p_2,q_2,p_3,q_3) =
\begin{array}{r}
{\epsfxsize 6cm\epsffile{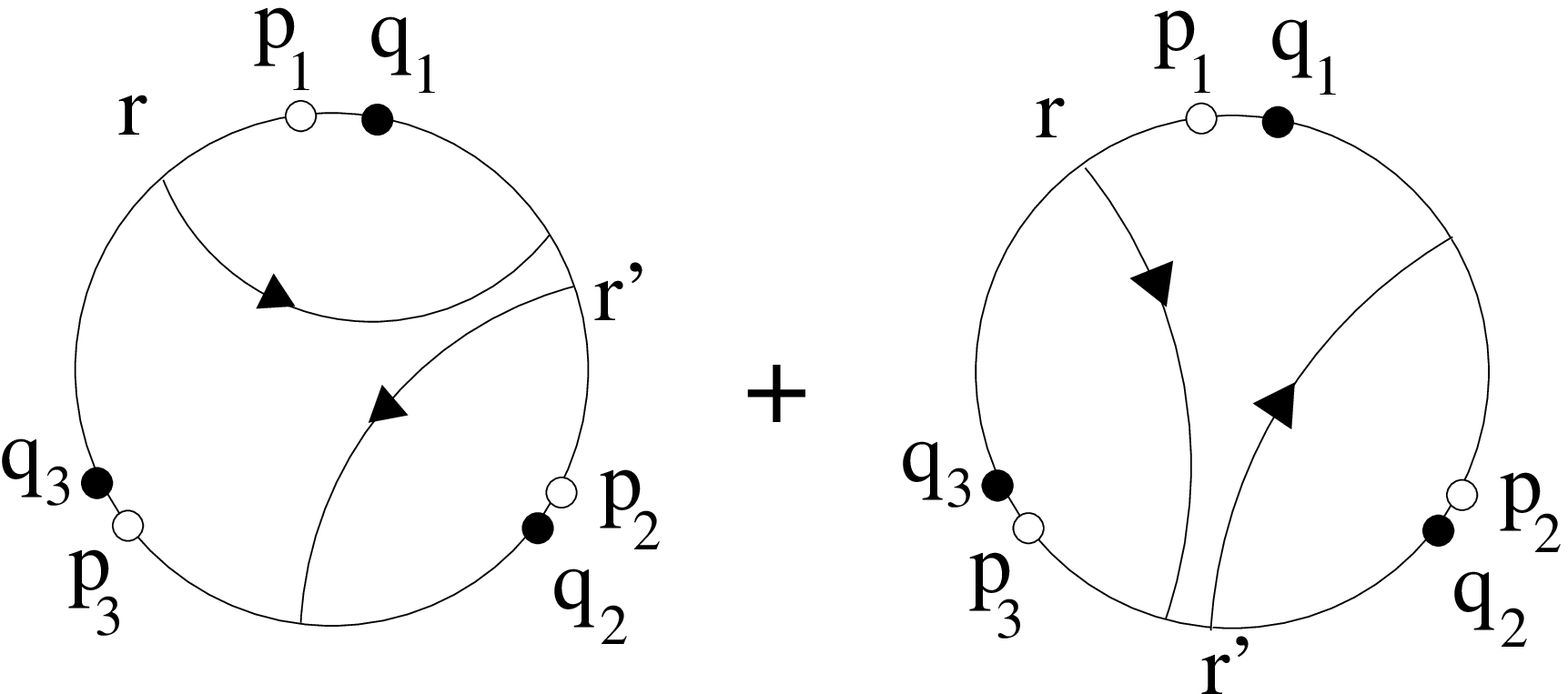}}
\end{array}
\eeq
which gives:
\beq
\begin{array}{l}
H_{3;0;0}^{(0)}(p_1,q_2,p_2,q_2,p_3,q_3) = \cr
 {\displaystyle \Res_{r \to p_1,p_2, \tilde{q}_3^j}  \Res_{r' \to p_2,p_3, \tilde{q}_3^j}}
{ H_{1;0;0}^{(0)}(p_1,q_3)  H_{1;0;0}^{(0)}(r,q_1)  H_{1;0;0}^{(0)}(p_2,q_3) H_{1;0;0}^{(0)}(p_3,q_3) H_{1;0;0}^{(0)}(r',q_2) \over
(x(p_1)-x(r))(y(q_3)-y(r))(x(p_2)-x(r)) H_{1;0;0}^{(0)}(r,q_3)(x(p_2)-x(r'))(y(q_3)-y(r'))(x(p_3)-x(r')) H_{1;0;0}^{(0)}(r',q_3)} \cr
+ {\displaystyle \Res_{r \to p_1,p_3, \tilde{q}_3^j}  \Res_{r' \to r,p_2, \tilde{q}_2^j}}
{ H_{1;0;0}^{(0)}(p_1,q_3)  H_{1;0;0}^{(0)}(r,q_2)  H_{1;0;0}^{(0)}(p_3,q_3) H_{1;0;0}^{(0)}(p_2,q_2) H_{1;0;0}^{(0)}(r',q_1) \over
(x(p_1)-x(r))(y(q_3)-y(r))(x(p_3)-x(r)) H_{1;0;0}^{(0)}(r,q_3)(x(r)-x(r'))(y(q_2)-y(r'))(x(p_2)-x(r')) H_{1;0;0}^{(0)}(r',q_2)} \cr
\end{array}
\eeq

One can easily show that this coincide with the result of \cite{EObethe} by using explicitly the orientation-reversing symmetry\footnote{Combinatoricaly, this means that summing over all oriented surfaces is equivalent to summing over all surfaces with the orientation reversed.} of the correlation function:
\beq
H_{3;0;0}^{(0)}(p_1,q_1,p_2,q_2,p_3,q_3) = H_{3;0;0}^{(0)}(p_1,q_3,p_3,q_2,p_2,q_1).
\eeq

\subsubsection{Simple traces topological expansion}

One can remark that all this recursive procedure supposes that the non-mixed correlation functions are known,
since this new diagrammatic representation does not allow to compute them. Nevertheless they were computed by
a similar procedure in \cite{eyno,CEO, EOFg} in terms of trivalent graphs and we show that these former rules could
be written in a graphical representation similar to the one presented in this paper.

Let us represent $W_{k+1}^{(h)}(p,p_1, \dots, p_k):= H_{0,k+1,0}^{(h)}(p,p_1, \dots, p_k)$ as a disk with
$k$ punctures instead of a sphere with $k+1$ punctures (we have drawn the surface generated by this function inside the boundary corresponding to $p$):
\beq
\begin{array}{r}
{\epsfxsize 3.4cm\epsffile{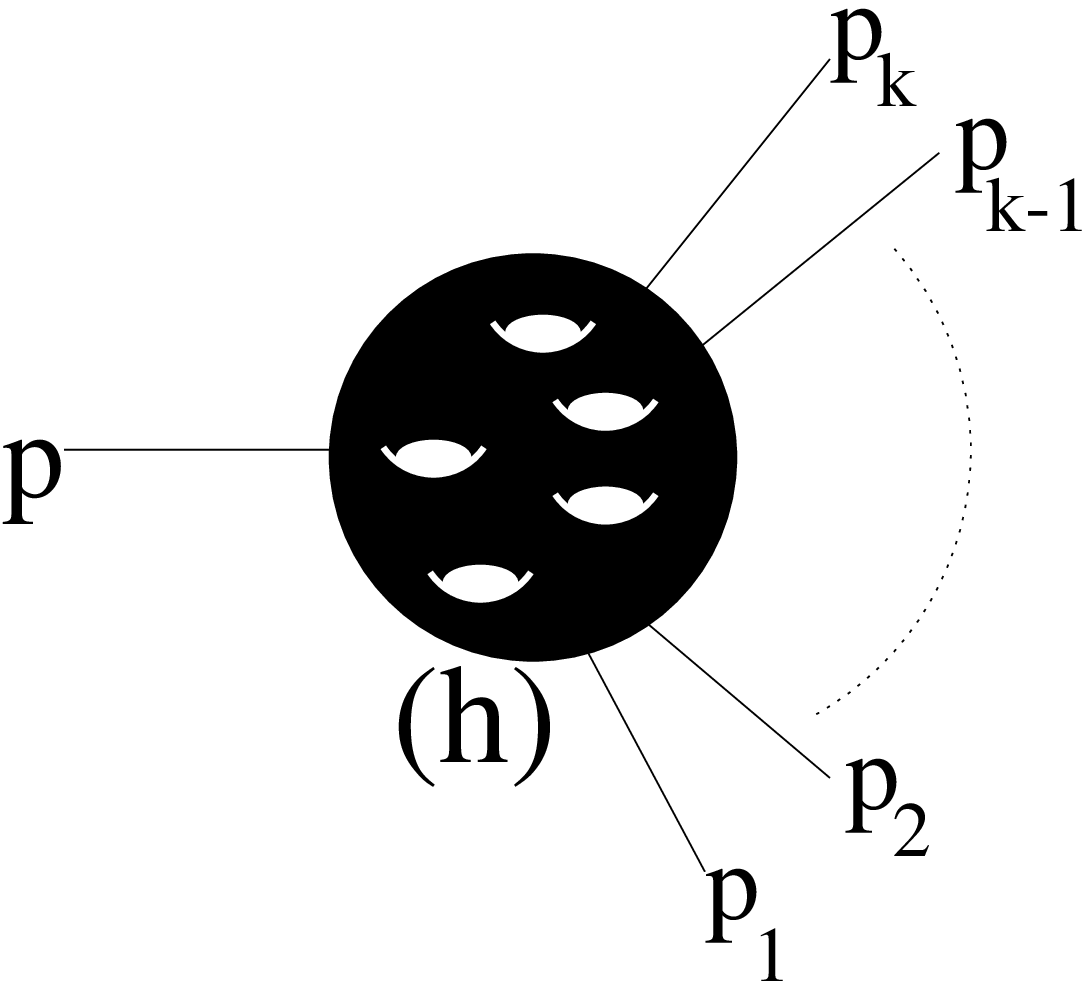}}
\end{array}
\Rightarrow
\begin{array}{r}
{\epsfxsize 2.8cm\epsffile{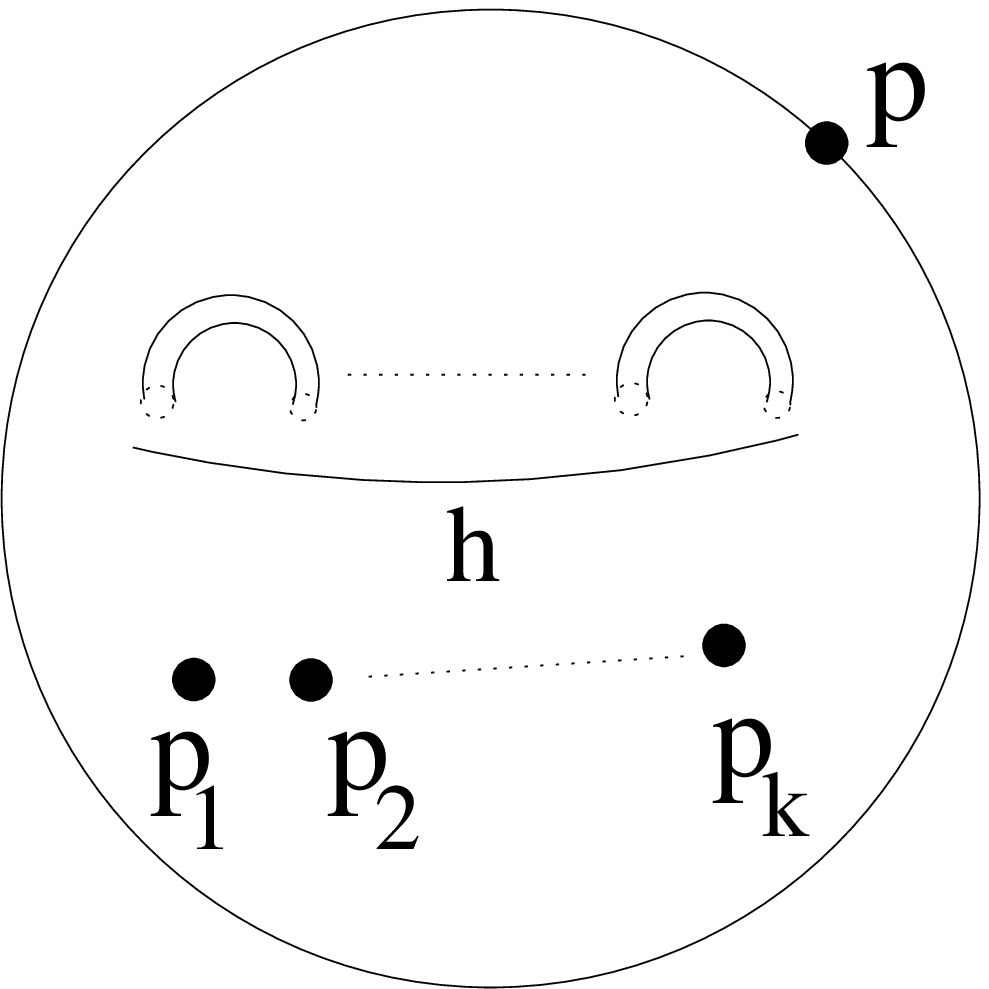}}
\end{array}
\eeq

The recursion relation of \cite{CEO, EOFg}
\beq
\begin{array}{r}
{\epsfxsize 3.4cm\epsffile{Wh.eps}}
\end{array}
=
\begin{array}{r}
{\epsfxsize 3.4cm\epsffile{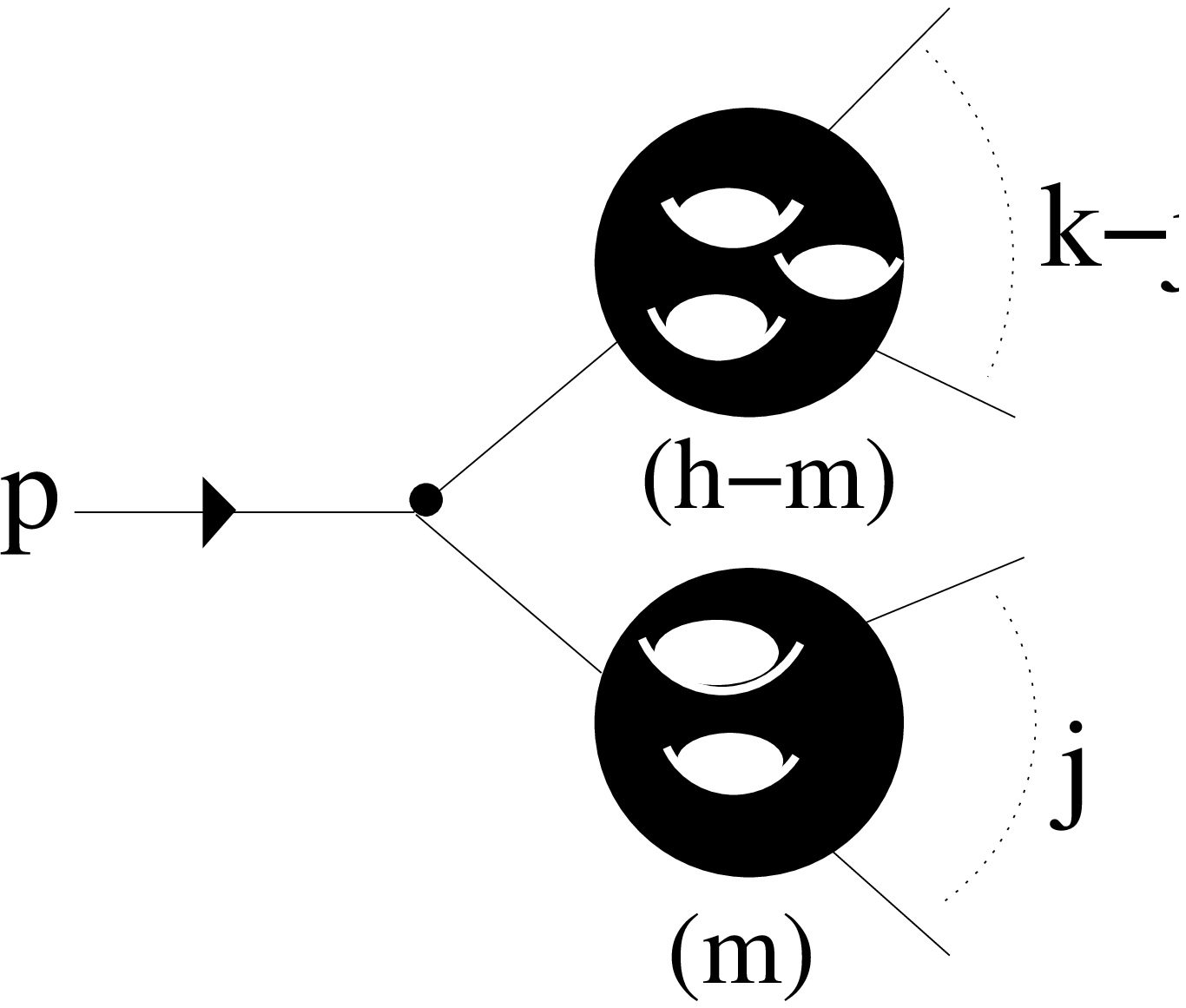}}
\end{array}
+
\begin{array}{r}
{\epsfxsize 3.4cm\epsffile{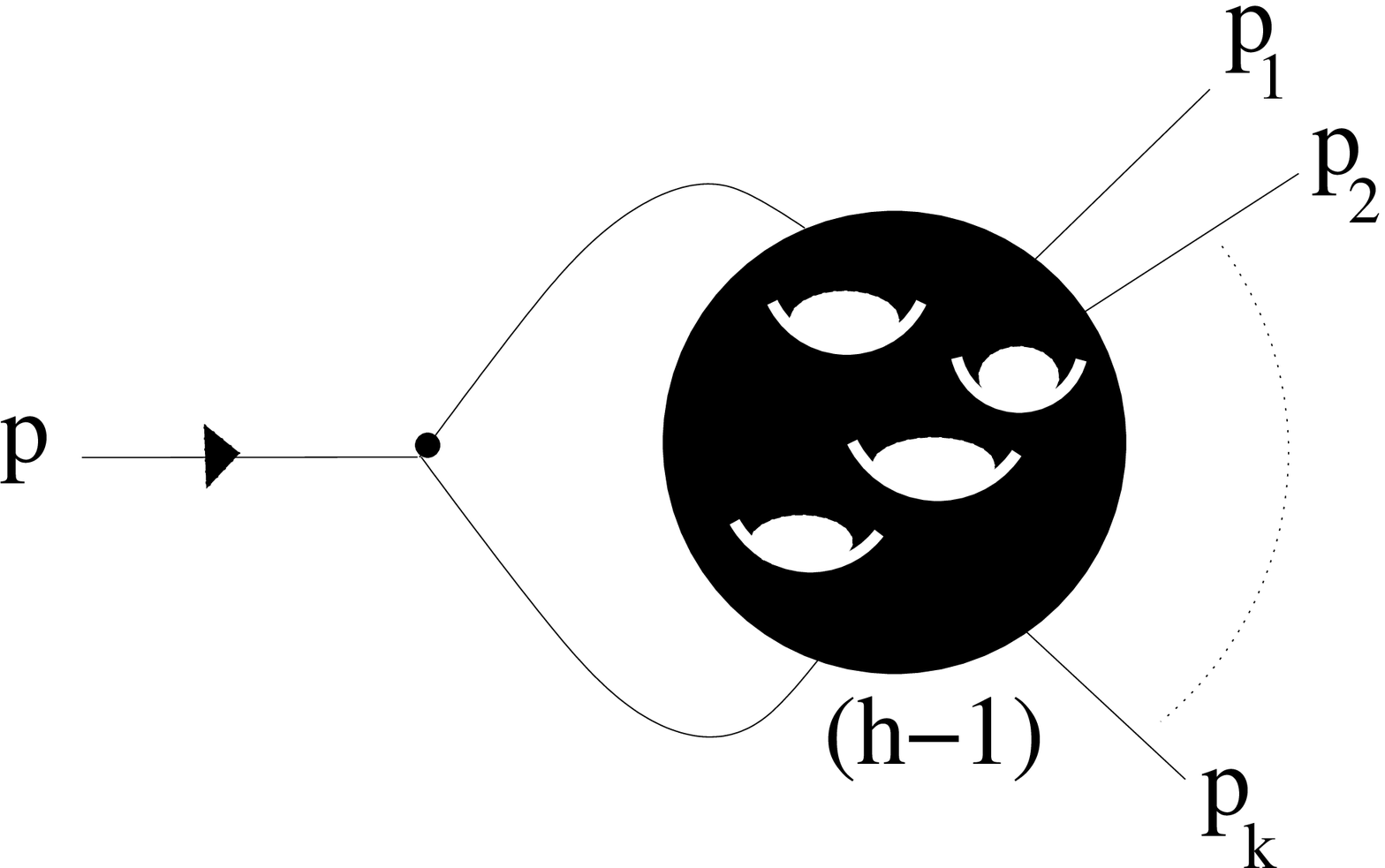}}
\end{array}
\eeq
can then be written:
\beq
\begin{array}{r}
{\epsfxsize 3.4cm\epsffile{simple.eps}}
\end{array}
=
\begin{array}{r}
{\epsfxsize 3.4cm\epsffile{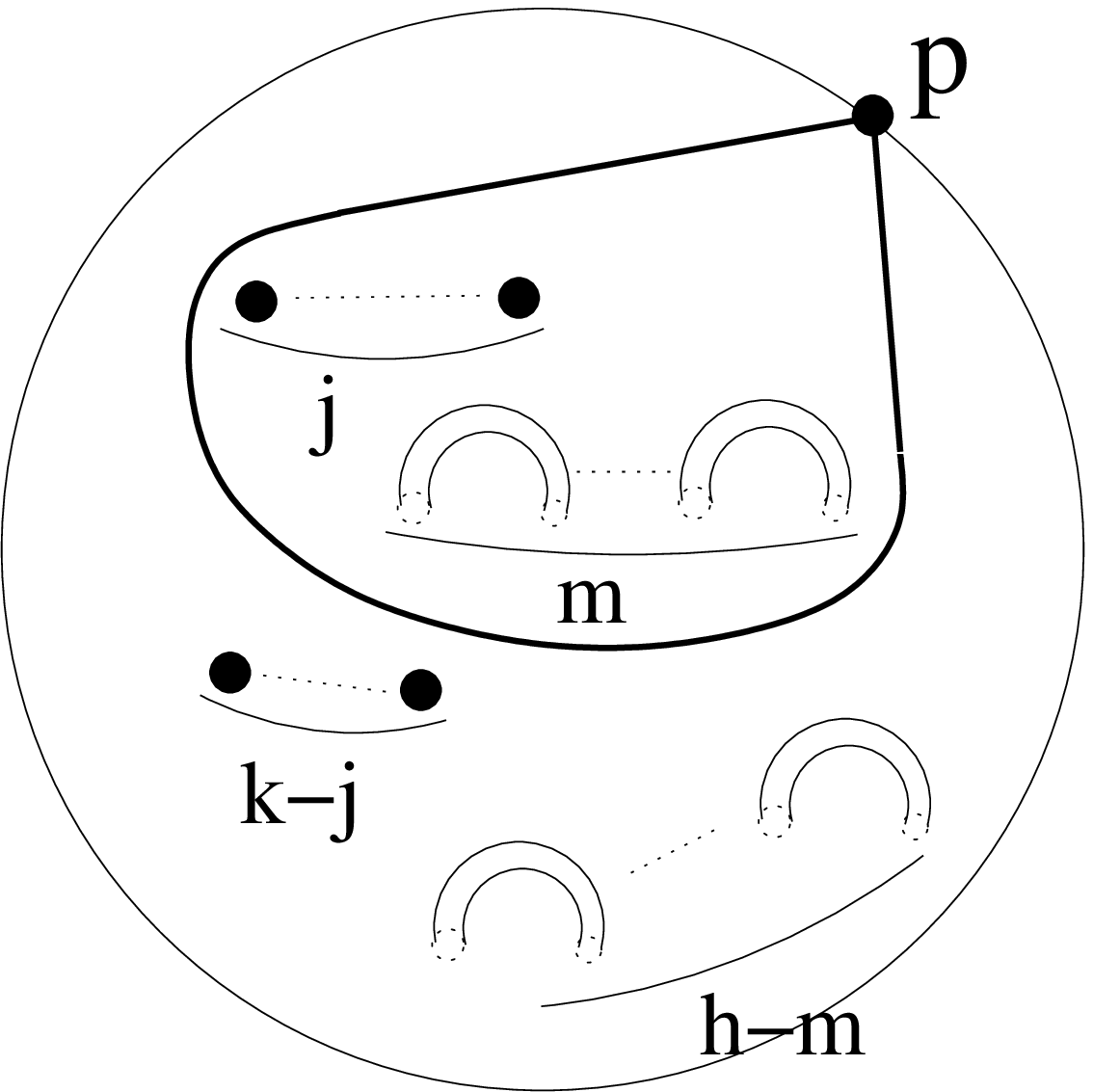}}
\end{array}
+
\begin{array}{r}
{\epsfxsize 3.4cm\epsffile{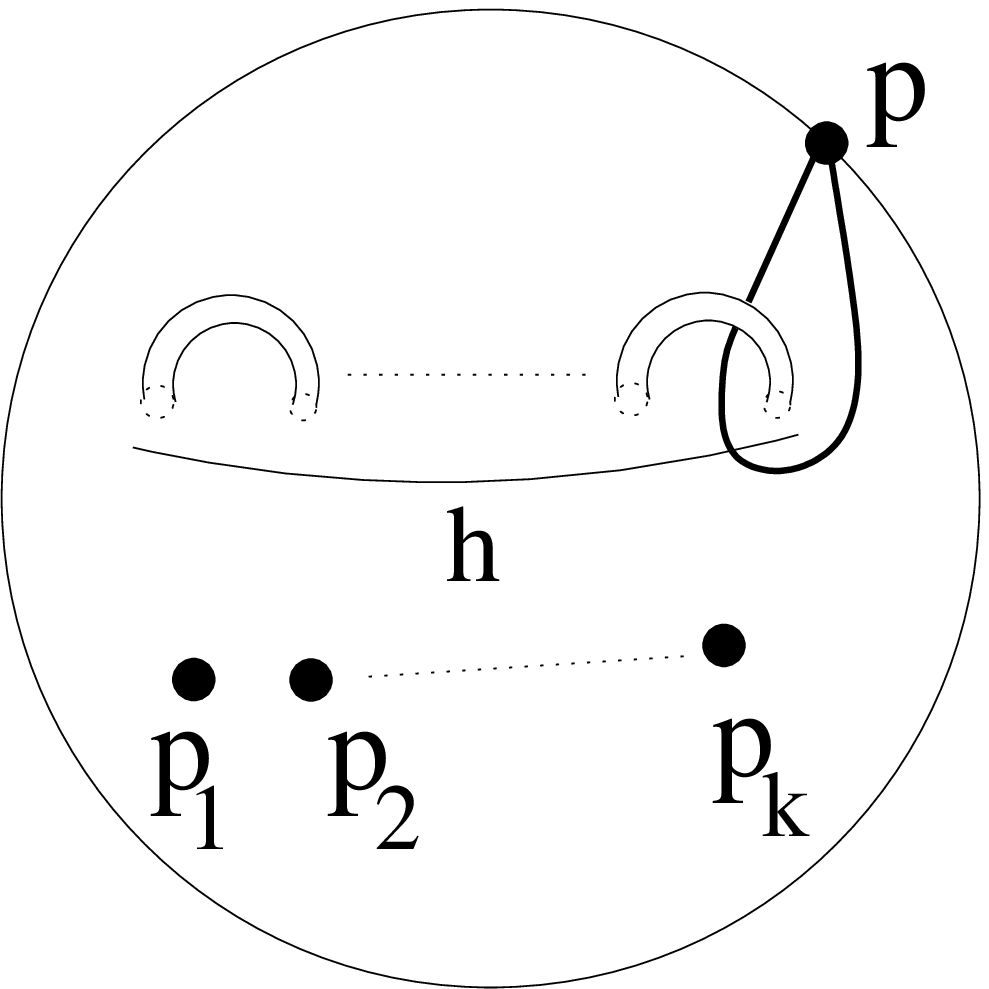}}
\end{array}
\eeq
where once again, the different terms on the RHS are obtained by drawing a basis of homologically independent
paths on the disk starting and ending on the boundary, and the weight of a cutting along this path follows:
\beq
\begin{array}{r}
{\epsfxsize 2.8cm\epsffile{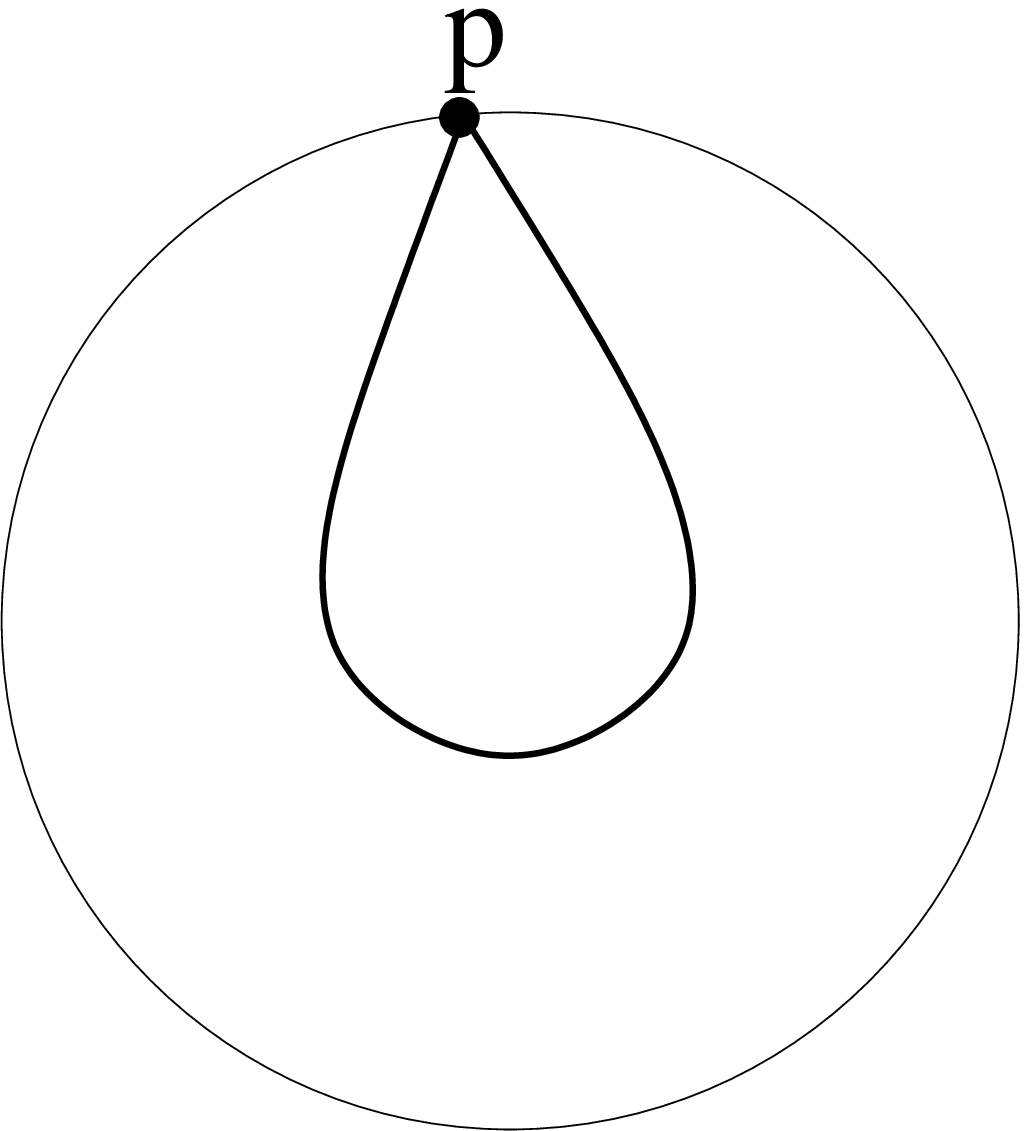}}
\end{array}
=
\sum_i \Res_{q \to a_i} {dE_q(p) \over (y(q)-y(\overline{q}))dx(q)}
\begin{array}{r}
{\epsfxsize 2.8cm\epsffile{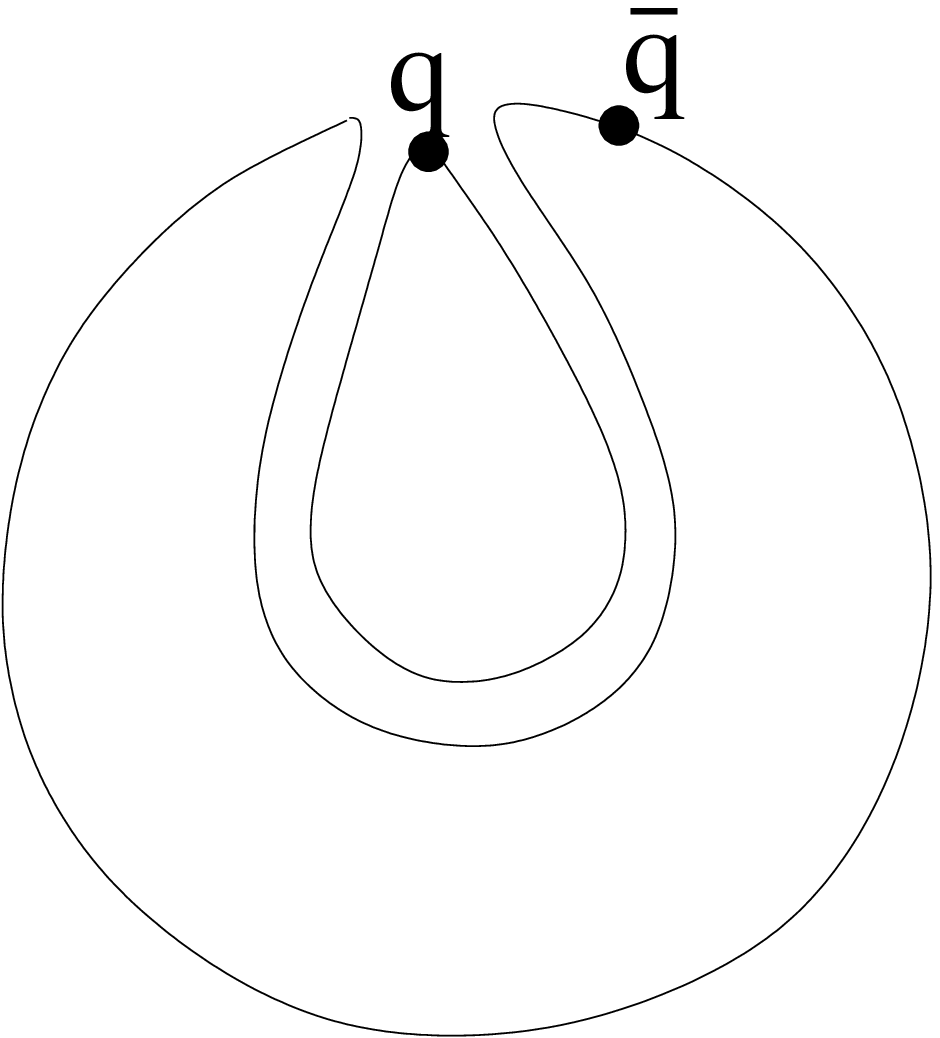}}
\end{array}
\eeq
where one sums over all branch points $a_i$ and $\overline{q}$ is the point conjugated to $q$ (see \cite{CEO, EOFg} for more details).

%\section{Examples}

\subsection{Four point function on the disc}

The correlation function $H_{2;0;0}^{(0)}(p_1,q_1,p_2,q_2)$ has already been computed in \cite{eyn2, EObethe}.
Nevertheless, this computation used an Ansatz and a symmetry property of the correlation function explicitly. Let us recover the same result without using any symmetry consideration, but using our recursive formula instead.

The solution of the loop equations reads graphically:
\beq
\begin{array}{r}
{\epsfxsize 2.5cm\epsffile{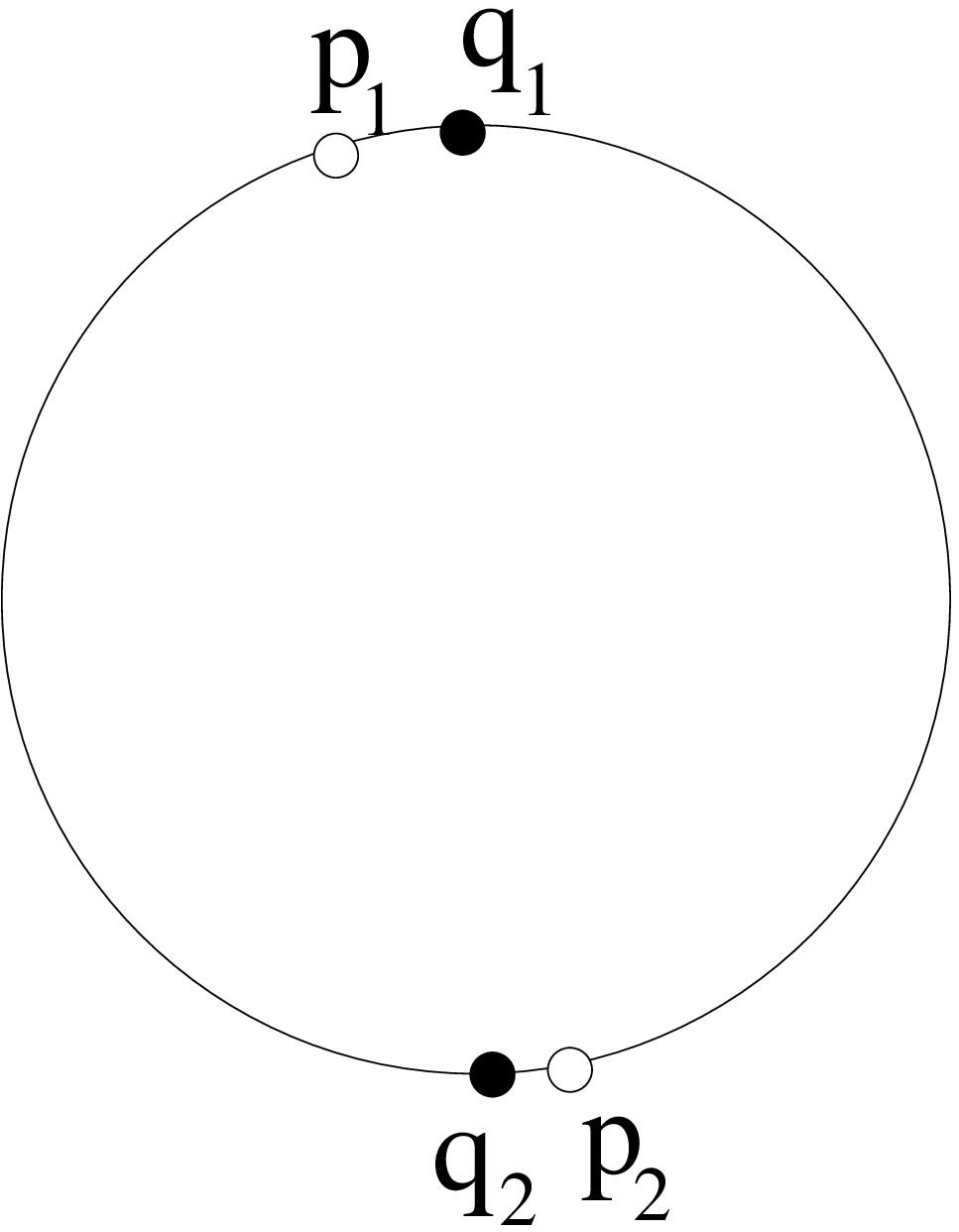}}
\end{array}
=
\begin{array}{r}
{\epsfxsize 2.5cm\epsffile{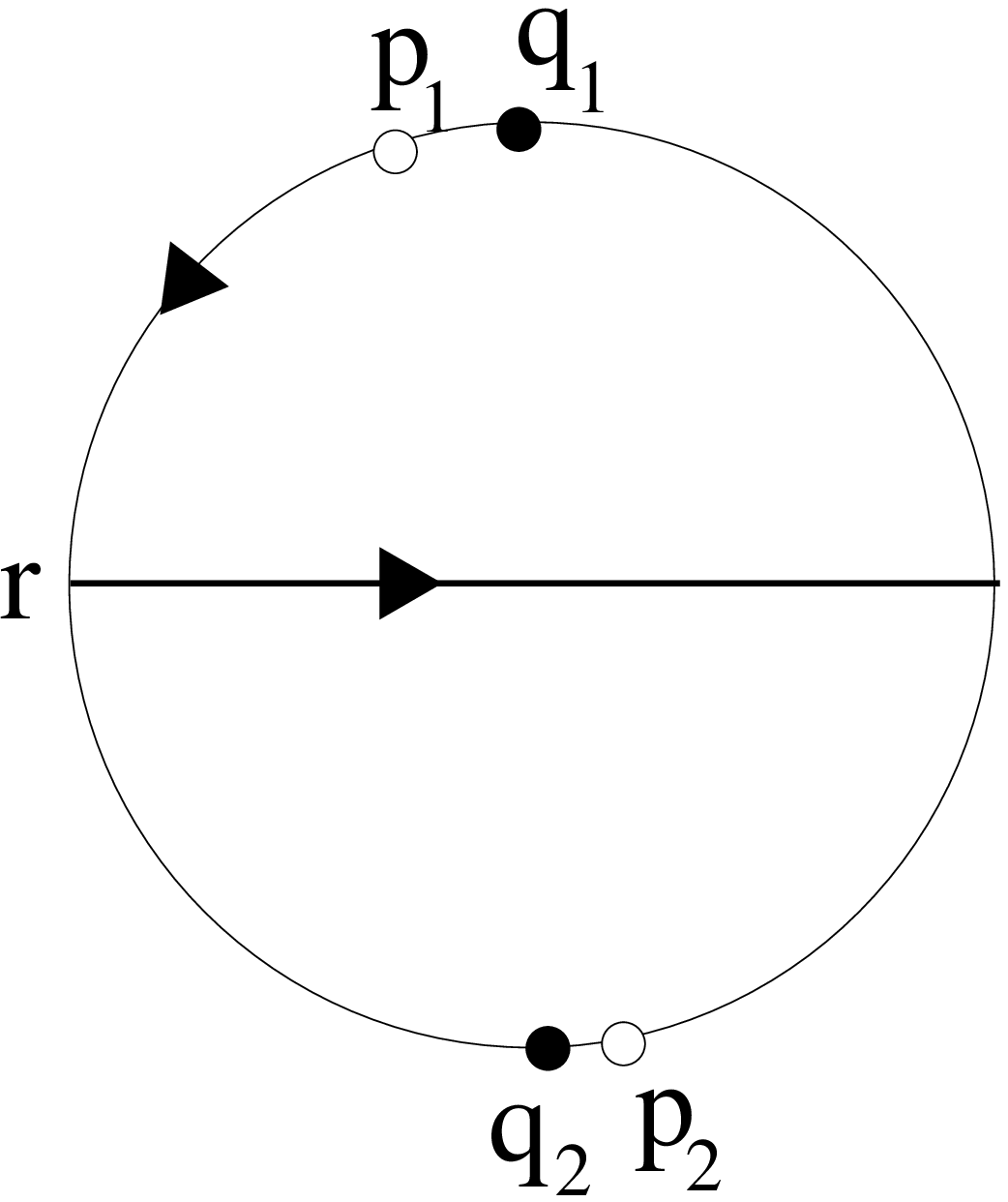}}
\end{array}
\eeq

which is translated into\footnote{For shortening the notations, we write all along this section
$H_{k}^{(g)}( p_1,q_1,p_2,q_2, \dots ,p_k,q_k):=H_{k;0;0}^{(g)}( p_1,q_1,p_2,q_2, \dots ,p_k,q_k)$.}
\beq
H_{2}^{(0)}(p_1,q_1,p_2,q_2) = \Res_{r \to p_1,p_2,\tilde{q}_2^j} {H_{1}^{(0)}(p_1,q_2)H_{1}^{(0)}(p_2,q_2) \,dx(r) \over H_{1}^{(0)}(r,q_2) (x(p_1)-x(r)) (y(q_2)-y(r)) } {H_{1}^{(0)}(r,q_1)\over x(p_2)-x(r)}.
\eeq
Writing
\beq
{1 \over y(q_2)-y(r)}= {y(r)-y(q_1) \over (y(q_2)-y(q_1)) (y(q_2)-y(r))}
+ {1 \over y(q_2)-y(q_1)}
\eeq
one gets
\bea
&& H_{2}^{(0)}(p_1,q_1,p_2,q_2) \cr
&=& \Res_{r \to p_1,p_2,\tilde{q}_2^j} {H_{1}^{(0)}(p_1,q_2)H_{1}^{(0)}(p_2,q_2) \, dx(r)\over H_{1}^{(0)}(r,q_2) (x(p_1)-x(r)) (y(q_2)-y(q_1)) } {H_{1}^{(0)}(r,q_1)\over x(p_2)-x(r)} \cr
&+& \Res_{r \to p_1,p_2,\tilde{q}_2^j} {H_{1}^{(0)}(p_1,q_2)H_{1}^{(0)}(p_2,q_2)\, dx(r) \over  (x(p_1)-x(r)) (x(p_2)-x(r)) (y(q_2)-y(q_1)) } {(y(r)-y(q_1)) H_{1}^{(0)}(r,q_1)\over (y(q_2)-y(r)) H_{1}^{(0)}(r,q_2)}.\cr
\eea
Since
\beq
H_{1}^{(0)}(p,q)(y(q)-y(p)) = {E(x(p),y(q)) \over x(p) - x(q)}
\eeq
the integrand of the second term in the RHS is a rational function in $x(r)$ and it is easily checked that the
integration contour encircles all its poles (this function is regular when $x(r) \to \infty$). Thus
this second term vanishes.

The first term has no pole at $r=\td{q}_2^j$, thus it involves only simple poles when $r \to p_1,p_2$ and we recover the known result
\cite{eyn2,EObethe}:
\beq
H_2^{(0)}(p_1,q_1,p_2,q_2) = -{H_1^{(0)}(p_1,q_1) H_1^{(0)}(p_2,q_2) - H_1^{(0)}(p_1,q_1) H_1^{(0)}(p_2,q_2)
\over (x(p_1)-x(p_2)) (y(q_1)-y(q_2))}.
\eeq

Even if this new derivation of an old result seems more involved technically, it has the advantage of being
constructive (the derivation of \cite{EObethe} was based on an ansatz) and does not suppose any additional symmetry of the correlation functions (the derivation of \cite{eyn2} was based on the fact that $H_2^{(0)}(p_1,q_1,p_2,q_2)=H_2^{(0)}(p_1,q_2,p_2,q_1)$).

\subsection{Generating function of cylinders}

The generating function of cylinders with 2 boundary operators on both boundaries is obtained by:
\beq
\begin{array}{r}
{\epsfxsize 2.5cm\epsffile{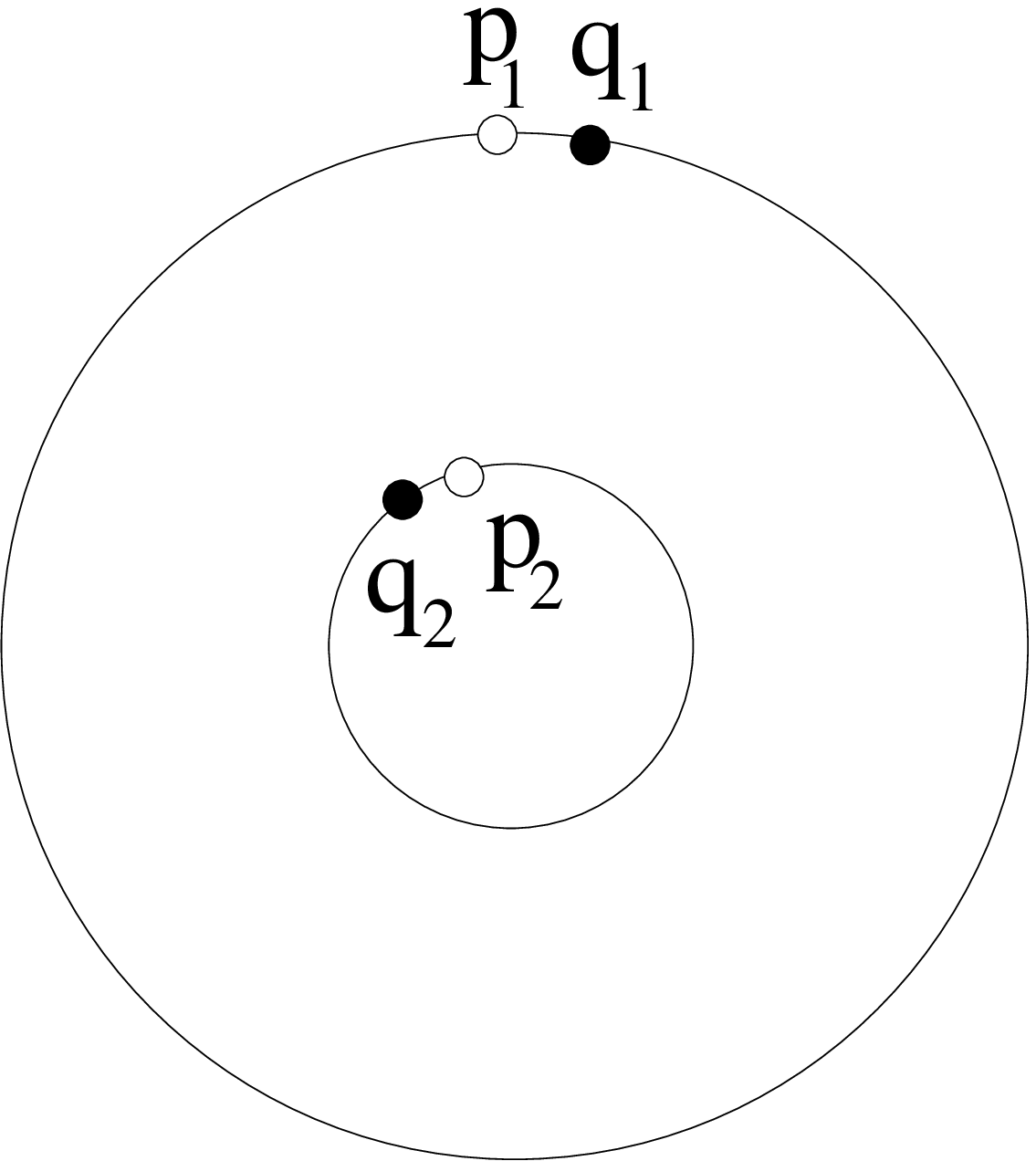}}
\end{array}
=
\begin{array}{r}
{\epsfxsize 2.5cm\epsffile{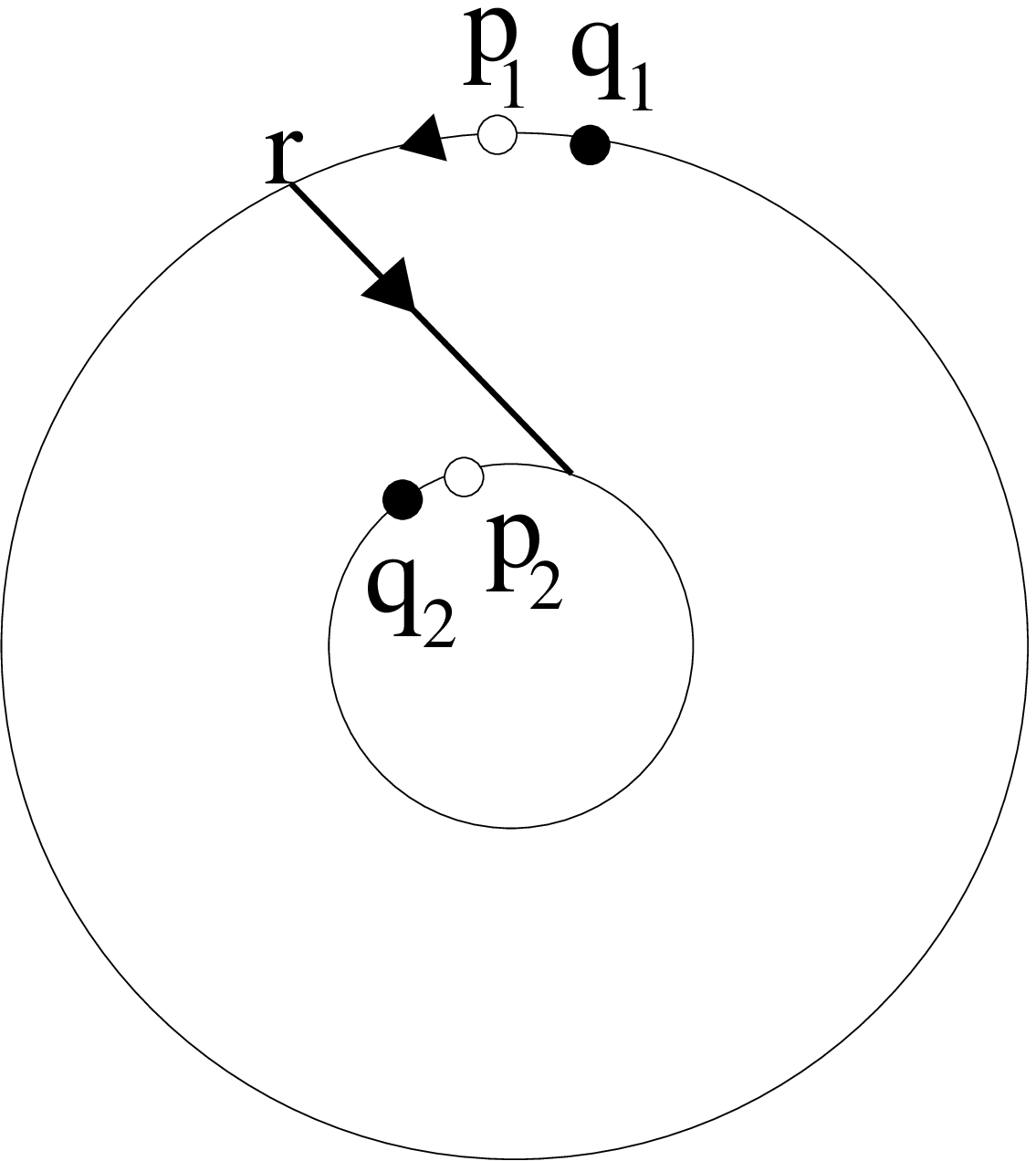}}
\end{array}
+
\begin{array}{r}
{\epsfxsize 2.5cm\epsffile{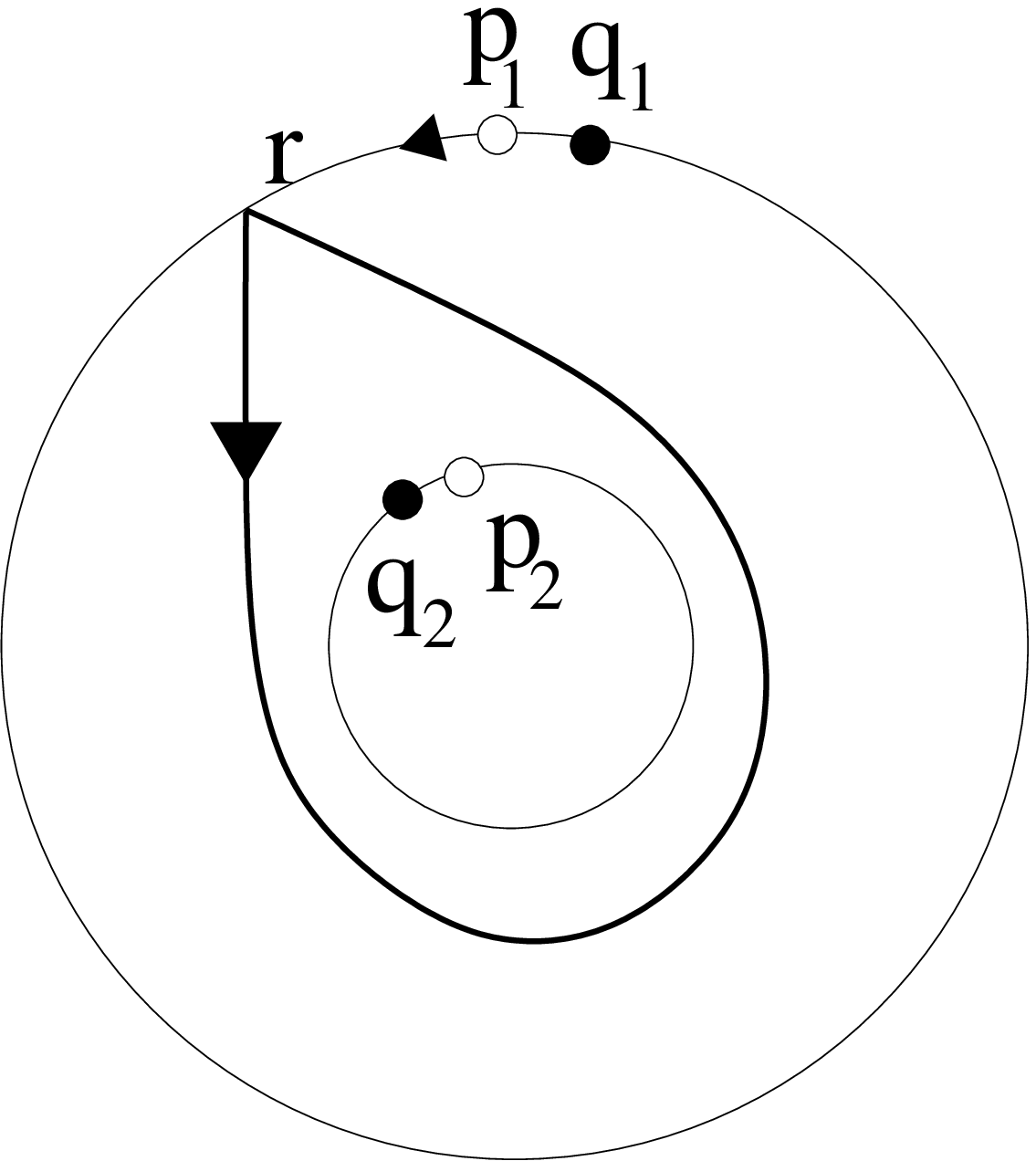}}
\end{array}
\eeq
which can be translated into
\bea
H_{1,1;0;0}^{(0)}(\{p_1,q_1\},\{p_2,q_2\})
&=& \Res_{r \to p_1,p_2,\tilde{q}_1^j} {H_{1}^{(0)}(p_1,q_1)\, dx(r) \over H_{1}^{(0)}(r,q_1) (x(p_1)-x(r)) (y(q_1)-y(r)) }
\times \cr
&& \;\;\;\; \times
\left[ {H_{2}^{(0)}(r,q_1,p_2,q_2) \over x(p_2)-x(r)} + H_{1}^{(0)}(r,q_1) H_{1;1;0}^{(0)}(p_2,q_2;r) \right]\cr
\eea
where the second term was computed in \cite{EOsymFg}:
\beq
H_{1;1;0}^{(0)}(p_2,q_2;r)= \Res_{r' \to p_2,r,\tilde{q_2}^j} {H_{1}^{(0)}(p_2,q_2) H_{0;2;0}^{(0)}(r;r')
\, dx(r) \over (x(p_2)-x(r')) (y(r')-y(q_2))}.
\eeq
and
$ H_{0;2;0}^{(0)}(r;r')\, dx(r) dx(r')$ is the Bergmann kernel.

\section{Conclusion}

In this article we have found a recursive and graphical method to compute correlation functions corresponding to every possible boundary condition for the 2-matrix model, i.e. bicolored discrete surfaces.

The result seems to have a nice combinatorial interpretation, as all the possibilities of drawing interfaces (between the + and - spins of the Ising model) in all possible ways.
However, a combinatorial derivation is missing.

Also, our result can have interpretations in conformal field theories when one goes to the so called double-scaling-limit \cite{ZJDFG, DKK}, and should be compared with recent results from Liouville theory \cite{ribault, schomerus, kostov}. In particular, in \cite{kostov}, our formula for planar disc amplitudes is interpreted in terms of the interactions of long folded strings and it would be interesting to check the non-planar cases as well.

It would be interesting also to understand how the structure we exhibit in this article, and which seems to be related to integrability like in \cite{EObethe}, is related to the Langlands programm as claimed by \cite{schomerus}.

\section*{Acknowledgments}
We would like to thank J.-E. Bourgine, L. Cantini, and I. Kostov  for useful and fruitful discussions on this subject.
This work is partly supported by the Enigma European network MRT-CT-2004-5652, by the ANR project G\'eom\'etrie et int\'egrabilit\'e en physique math\'ematique ANR-05-BLAN-0029-01, by the Enrage European network MRTN-CT-2004-005616,
by the European Science Foundation through the Misgam program,
by the French and Japanese governments through PAI Sakurav, by the Quebec government with the FQRNT, and the Centre de Recherche Math\'ematiques de Montr\'eal, QC, Canada.

\setcounter{section}{0}
\appendix{Loop equations}\label{proofloopeq}

Here we prove the equation \ref{soluce}, using loop equations.
Loop equations is a standard and powerful tool in random matrix theory, they are the Ward identities, or Schwinger-Dyson equations, they implement the Virasoro or W-algebra constraints, in combinatorics they can be viewed as an extension of Tutte's equations, and in fact they just consist in integration by parts, or said differentely, the fact that an integral is invariant under (an infinitesimal) change of variable.

For the 2-matrix model, loop equations were first exploited by Staudacher \cite{staudacher}, and then by many authors, and they led to the solution of \cite{eyno, CEO, EOsymFg}.

\subsection{The loop equations}

In order to prove eq.\ref{soluce}, we consider the change of variables
\beq
\begin{array}{rcl}
\delta M_1 &:=&{1 \over x(p_{1,1})-M_1}{1 \over y(p_{1,1})-M_2} {1 \over x(p_{1,2})-M_1}{1 \over y(p_{1,2})-M_2}
\dots {1 \over x(p_{1,k_1})-M_1}{1 \over y(p_{1,k_1})-M_2} \cr
&&
\prod_{i=2}^l \Tr \left({1 \over x(p_{i,1})-M_1}{1 \over y(p_{i,1})-M_2} {1 \over x(p_{i,2})-M_1}{1 \over y(p_{i,2})-M_2}
\dots {1 \over x(p_{i,k_i})-M_1}{1 \over y(p_{i,k_i})-M_2}\right) \cr
&& \qquad \; \; \prod_{j=1}^m \Tr { 1 \over x(p_j)-M_1} \prod_{s=1}^n \Tr {1 \over y(q_s)-M_2}.\cr
\end{array}
\eeq

Writing that the matrix integral is invariant under this change of variable gives the loop equation:
\beq
\begin{array}{l}
(Y(p_{1,1})-y(q_{1,k_1})-  Pol_{x(p_{1,1})} V_1'(x(p_{1,1})) ) H_{k_1,\dots,k_l;m;n}(S_1,S_2, \dots, S_l;p_1 ,\dots, p_m;q_1,\dots,q_n) = \cr
%%%%%%%%%%%%%%%%
%%%%%%%%%%%%%%%%
\sum_{A\bigcup B=\{2, \dots,l\}}  \sum_{I,J} H_{k_1,{\bf k_A};\left|I\right|;\left|J\right|}(S_1,{\bf S_A}; {\bf p_{I}};{\bf q_J})
H_{{\bf k_B};m-\left|I\right|+1;n-\left|J\right|}({\bf S_B}; p_{1,1} {\bf p_{M/I}};{\bf q_N/J})\cr
%%%%%%%%%%%%%%%%
%%%%%%%%%%%%%%%%
+ \sum_{A\bigcup B=\{2, \dots,l\}}\sum_{\alpha =2}^{k_1} \sum_{I,J} H_{k_1-\alpha+1,{\bf k_B};m-\left|I\right|;n-\left|J\right|}(\{p_{1,\alpha},q_{1,\alpha},\dots p_{1,k_1},q_{1,k_1}\},{\bf S_B}; {\bf p_{M/I}};{\bf q_{N/J}})\cr
\qquad  \times { H_{\alpha-1,{\bf k_A};\left|I\right|;\left|J\right|}(\{p_{1,1},q_{1,1},\dots p_{1,\alpha-1},q_{1,\alpha-1}\},{\bf S_A}; {\bf p_{I}};{\bf q_J})
-  H_{\alpha-1,{\bf k_A};\left|I\right|;\left|J\right|}(\{p_{1,\alpha},q_{1,1},\dots p_{1,\alpha-1},q_{1,\alpha-1}\},{\bf S_A}; {\bf p_{I}};{\bf q_J}) \over x(p_{1,\alpha})-x(p_{1,1})} \cr
%%%%%%%%%%%%%%%%
%%%%%%%%%%%%%%%%
\sum_{i=2}^{l} \sum_{\alpha=1}^{k_i} {1 \over x(p_{i,\alpha}) - x(p_{1,1})} \cr
\left[ H_{k_1+k_i,{\bf k_{L/\{1,i\}}};m;n}( \{S_1,p_{i,\alpha},q_{i,\alpha},p_{i,\alpha+1}, \dots ,q_{i,k_i},p_{i,1},\dots ,p_{i,\alpha-1},q_{i,\alpha-1}\},{\bf S_{L/\{1,i\}}};{\bf p_M};{\bf q_N}) \right. \cr
\left. - \left. H_{k_1+k_i,{\bf k_{L/\{1,i\}}};m;n}( \{S_1,p_{i,\alpha},q_{i,\alpha},p_{i,\alpha+1}, \dots ,q_{i,k_i},p_{i,1},\dots ,p_{i,\alpha-1},q_{i,\alpha-1}\},{\bf S_{L/\{1,i\}}};{\bf p_M};{\bf q_N})\right|_{p_{1,1}:=p_{i,\alpha}} \right] \cr
%%%%%%%%%%%%%%%%
%%%%%%%%%%%%%%%%
- \sum_{i=1}^{m} \partial_{p_i} \left[{H_{{\bf k_L};m-1;n}({\bf S_L};{\bf p_{M/\{i\}}};{\bf q_N}) - \left. H_{{\bf k_L};m-1;n}({\bf S_L};{\bf p_{M/\{i\}}};{\bf q_N})\right|_{p_{1,1}:=p_i} \over x(p_i) - x(p_{1,1})} \right] \cr
%%%%%%%%%%%%%%%%
%%%%%%%%%%%%%%%%
+ {1 \over N^2} \sum_{\alpha =2}^{k_1} { 1 \over x(p_{1,\alpha})-x(p_{1,1})} \times \cr
\left[ H_{\alpha-1, k_1-\alpha+1, { \bf k_{L/\{1\}}}; m;n}(\{p_{1,1},q_{1,1},\dots p_{1,\alpha-1},q_{1,\alpha-1}\},\{p_{1,\alpha},q_{1,\alpha},\dots p_{1,k_1},q_{1,k_1}\}, {\bf S_{L/\{1\}}};{\bf p_M};{\bf q_N}) \right. \cr
\left. - H_{\alpha-1, k_1-\alpha+1, { \bf k_{L/\{1\}}}; m;n}(\{p_{1,\alpha},q_{1,1},\dots p_{1,\alpha-1},q_{1,\alpha-1}\},\{p_{1,\alpha},q_{1,\alpha},\dots p_{1,k_1},q_{1,k_1}\}, {\bf S_{L/\{1\}}};{\bf p_M};{\bf q_N}) \right] \cr
%%%%%%%%%%%%%%%%
%%%%%%%%%%%%%%%%
+ {1 \over N^2} H_{{\bf k_L};m+1;n}({\bf S_K};p_{1,1},{\bf p_M};{\bf q_N})\cr
\end{array}
\eeq
where $Pol_x f(x)$ denotes the polynomial part in $x$ of $f$, i.e. the sum of the positive terms in the large $x$  Laurent expansion of $f(x)$.

Let us now write its $g^{\rm th}$ order in the topological expansion:
\beq \label{loopeq}
\begin{array}{l}
(y(p_{1,1})-y(q_{1,k_1})-  Pol_{x(p_{1,1})} V_1'(x(p_{1,1})) ) H_{k_1,\dots,k_l;m;n}^{(g)}(S_1,S_2, \dots, S_l;p_1 ,\dots, p_m;q_1,\dots,q_n) = \cr
%%%%%%%%%%%%%%%%
%%%%%%%%%%%%%%%%
\sum_{h=1}^{g} H_{0;1;0}^{(h)}(p_{1,1}) H_{k_1,\dots,k_l;m;n}^{(g-h)}(S_1,S_2, \dots, S_l;p_1 ,\dots, p_m;q_1,\dots,q_n) \cr
%%%%%%%%%%%%%%%%
%%%%%%%%%%%%%%%%
+\sum_h \sum_{A\bigcup B=\{2, \dots,l\}}  \sum_{I,J} H_{k_1,{\bf k_A};\left|I\right|;\left|J\right|}^{(h)}(S_1,{\bf S_A}; {\bf p_{I}};{\bf q_J})
H_{{\bf k_B};m-\left|I\right|+1;n-\left|J\right|}^{(g-h)}({\bf S_B}; p_{1,1} {\bf p_{M/I}};{\bf q_N/J})\cr
%%%%%%%%%%%%%%%%
%%%%%%%%%%%%%%%%
+ \sum_h \sum_{A\bigcup B=\{2, \dots,l\}}\sum_{\alpha =2}^{k_1} \sum_{I,J} H_{k_1-\alpha+1,{\bf k_B};m-\left|I\right|;n-\left|J\right|}^{(h)}(\{p_{1,\alpha},q_{1,\alpha},\dots p_{1,k_1},q_{1,k_1}\},{\bf S_B}; {\bf p_{M/I}};{\bf q_{N/J}})\cr
\qquad  \times { H_{\alpha-1,{\bf k_A};\left|I\right|;\left|J\right|}^{(g-h)}(\{p_{1,1},q_{1,1},\dots p_{1,\alpha-1},q_{1,\alpha-1}\},{\bf S_A}; {\bf p_{I}};{\bf q_J})
-  H_{\alpha-1,{\bf k_A};\left|I\right|;\left|J\right|}^{(g-h)}(\{p_{1,\alpha},q_{1,1},\dots p_{1,\alpha-1},q_{1,\alpha-1}\},{\bf S_A}; {\bf p_{I}};{\bf q_J}) \over x(p_{1,\alpha})-x(p_{1,1})} \cr
%%%%%%%%%%%%%%%%
%%%%%%%%%%%%%%%%
+ \sum_{i=2}^{l} \sum_{\alpha=1}^{k_i} {1 \over x(p_{i,\alpha}) - x(p_{1,1})} \cr
\left[ H_{k_1+k_i,{\bf k_{L/\{1,i\}}};m;n}^{(g)}( \{S_1,p_{i,\alpha},q_{i,\alpha},p_{i,\alpha+1}, \dots ,q_{i,k_i},p_{i,1},\dots ,p_{i,\alpha-1},q_{i,\alpha-1}\},{\bf S_{L/\{1,i\}}};{\bf p_M};{\bf q_N}) \right. \cr
\left. - \left. H_{k_1+k_i,{\bf k_{L/\{1,i\}}};m;n}^{(g)}( \{S_1,p_{i,\alpha},q_{i,\alpha},p_{i,\alpha+1}, \dots ,q_{i,k_i},p_{i,1},\dots ,p_{i,\alpha-1},q_{i,\alpha-1}\},{\bf S_{L/\{1,i\}}};{\bf p_M};{\bf q_N})\right|_{p_{1,1}:=p_{i,\alpha}} \right] \cr
%%%%%%%%%%%%%%%%
%%%%%%%%%%%%%%%%
+ \sum_{i=1}^{m} \partial_{p_i} \left[{\left. H_{{\bf k_L};m-1;n}({\bf S_L};{\bf p_{M/\{i\}}};{\bf q_N})\right|_{p_{1,1}:=p_i} \over x(p_i) - x(p_{1,1})} \right] \cr
%%%%%%%%%%%%%%%%
%%%%%%%%%%%%%%%%
+ \sum_{\alpha =2}^{k_1} { 1 \over x(p_{1,\alpha})-x(p_{1,1})} \times \cr
\left[ H_{\alpha-1, k_1-\alpha+1, { \bf k_{L/\{1\}}}; m;n}^{(g-1)}(\{p_{1,1},q_{1,1},\dots p_{1,\alpha-1},q_{1,\alpha-1}\},\{p_{1,\alpha},q_{1,\alpha},\dots p_{1,k_1},q_{1,k_1}\}, {\bf S_{L/\{1\}}};{\bf p_M};{\bf q_N}) \right. \cr
\left. - H_{\alpha-1, k_1-\alpha+1, { \bf k_{L/\{1\}}}; m;n}^{(g-1)}(\{p_{1,\alpha},q_{1,1},\dots p_{1,\alpha-1},q_{1,\alpha-1}\},\{p_{1,\alpha},q_{1,\alpha},\dots p_{1,k_1},q_{1,k_1}\}, {\bf S_{L/\{1\}}};{\bf p_M};{\bf q_N}) \right] \cr
%%%%%%%%%%%%%%%%
%%%%%%%%%%%%%%%%
+ H_{{\bf k_L};m+1;n}^{(g-1)}({\bf S_K};p_{1,1},{\bf p_M};{\bf q_N}).\cr
\end{array}
\eeq
Notice that we have used the normalizations $H_{0;2;0} = \left< \Tr \Tr \right>_c + {1 \over (x-x)^2}$ explicitly.

\subsection{Solution of the equations}

We can solve this hierarchy of equations by induction in the number of traces in the correlations and the genus.
Indeed, one can remark that the RHS of \eq{loopeq} contains correlation functions with either less traces (that is to say less
arguments) either lower genus compare to the correlation function in the LHS. One also knows that the last term of
the LHS is a polynomial in $x(p_{1,1})$ of degree $d_1-1$ and one can compute its value in the $d_1$ points $p_{1,1}\to \tilde{q}_{1,k_1}^{j}$
for $j = 1 \dots d_1$ independently of $H_{{\bf k_L};m;n}^{(g)}({\bf S_K};{\bf p_M};{\bf q_N})$.

For this purpose, let us study the behavior of the LHS when $p_{1,1} \to \tilde{q}_{1,k_1}^j$. If $p$ lies in the $x$-physical sheet
and $q_{1,k_1}$ to the $y$-physical sheet, the definition of the correlation function reads:
\bea
&&(y(p_{1,1})-y(q_{1,k_1})) H_{k_1,\dots,k_l;m;n}^{(g)}(S_1,S_2, \dots, S_l;p_1 ,\dots, p_m;q_1,\dots,q_n)
= \cr
&& = (y(p_{1,1})-y(q_{1,k_1})) \Big<
\prod_{i=1}^l  (N \delta_{k_i,1}+\Tr {1\over S_i})  \,\,
 \prod_{j=1}^m \Tr { 1 \over x(p_j)-M_1} \prod_{s=1}^n \Tr {1 \over y(q_s)-M_2}
\Big>_c^{(g)}\cr
&& = - \Big< (N \delta_{k_i,1}+\Tr {1\over \widehat{S}_1})
\prod_{i=2}^l  (N \delta_{k_i,1}+\Tr {1\over S_i})  \,\,
 \prod_{j=1}^m \Tr { 1 \over x(p_j)-M_1} \prod_{s=1}^n \Tr {1 \over y(q_s)-M_2}
\Big>_c^{(g)}\cr
&& \; \;  + \Big< (N \delta_{k_i,1}+\Tr {1\over \check{S}_1})
\prod_{i=2}^l  (N \delta_{k_i,1}+\Tr {1\over S_i})  \,\,
 \prod_{j=1}^m \Tr { 1 \over x(p_j)-M_1} \prod_{s=1}^n \Tr {1 \over y(q_s)-M_2}
\Big>_c^{(g)}\cr
\eea
where one notes:
\beq
\Tr {1\over \widehat{S}_i} = \Tr \left({1 \over x(p_{1,1})-M_1}{1 \over y(q_{1,1})-M_2}
{1 \over x(p_{1,2})-M_1}
{1 \over y(q_{1,2})-M_2}
\dots
{1 \over x(p_{1,k_1})-M_1}\right)
\eeq
and
\beq
\Tr {1\over \widehat{S}_i} = \Tr \left({1 \over x(p_{1,1})-M_1}{1 \over y(q_{1,1})-M_2}
{1 \over x(p_{1,2})-M_1}
{1 \over y(q_{1,2})-M_2}
\dots
{1 \over x(p_{1,k_1})-M_1}
{y(p_{1,1})-M_2 \over y(q_{1,k_1})-M_2}\right).
\eeq
These terms are monovalued functions as long as the $p$ and $q$ variables stay in their respective physical sheets. When
$q_{1,k_1}$ belongs to the $y$-physical sheet in the vincinity of $\infty_y$, all its images $\tilde{q}_{1,k_1}^{j}$ lie
in the $x$-physical sheet in the vincinity of $\infty_x$. Thus, this expression vanishes for $p_{1,1} \to \tilde{q}_{1,k_1}^j$\footnote{
This term does not vanish when $p_{1,1} \to q_{1,k_1}$ because of the discontinuity of these functions when $p_{1,1}$ changes
$x$-sheets.}.
Hence the Lagrange interpolation formula reads
\beq
U_{k_1,\dots,k_l;m;n}^{(g)}(x(p_{1,1}))=
\sum_{j=1}^{d_1} \Res_{r \to \tilde{q}^{j}}{ H_{1;0;0}^{(0)}(p_{1,1},q_{1,k_1})U_{k_1,\dots,k_l;m;n}^{(g)}(x(r))(y(p_{1,1})-y(q)) dx(r) \over
(x(p_{1,1})-x(r))(y(r)-y(q))H_{1;0;0}^{(0)}(r,q_{1,k_1})},
\eeq
where we have defined:
\beq
U_{k_1,\dots,k_l;m;n}^{(g)}(x(p_{1,1})):=
Pol_{x(p_{1,1})} V_1'(x(p_{1,1}))  H_{k_1,\dots,k_l;m;n}^{(g)}(S_1,S_2, \dots, S_l;p_1 ,\dots, p_m;q_1,\dots,q_n).
\eeq

Insert this formula into \eq{loopeq} and get:
\beq
H_{{\bf k_L} ;m;n}^{(g)}({\bf S_L};p_1 ,\dots, p_m;q_1,\dots,q_n) =
\Res_{r \to p_{1,1} , \tilde{q}^{j}} { H_{1;0;0}^{(0)}(p_{1,1},q_{1,k_1})\; \hbox{RHS}|_{p_{1,1}:=r} \over
(x(p_{1,1})-x(r))(y(q)-y(r))H_{1;0;0}^{(0)}(r,q_{1,k_1})}
\eeq
where RHS denotes all the terms in the right hand side of \eq{loopeq}.

One can simplify some of the terms by changing the integration contour. Indeed, consider any term of the form
$d_{p_{i,\alpha}} \left({f(p_{i,\alpha}) \over x(p_{1,1}) - x(p_{i,\alpha})} \right)$ in the RHS of \eq{loopeq}, one can compute
its contribution to the preceding formula:
\beq
\begin{array}{l}
d_{p_{i,\alpha}}\Res_{r \to p_{1,1} , \tilde{q}^{j}} { H_{1;0;0}^{(0)}(p_{1,1},q_{1,k_1})f(p_{i,\alpha}) \over
(x(p_{1,1})-x(r))(y(q)-y(r))(x(p_{1,1}) - x(p_{i,\alpha}))H_{1;0;0}^{(0)}(r,q_{1,k_1})} =\cr
=d_{p_{i,\alpha}}\Res_{x(r) \to x(p_{1,1}) , x(\tilde{q}^{j})} { H_{1;0;0}^{(0)}(p_{1,1},q_{1,k_1})f(p_{i,\alpha}) \over
(x(p_{1,1})-x(r))(y(q)-y(r))(x(p_{1,1}) - x(p_{i,\alpha}))H_{1;0;0}^{(0)}(r,q_{1,k_1})} \cr
\end{array}
\eeq
since one can check that the integrand is a polynomial in $x(r)$. We can now move the integration contour in the $x$
basis and we get:
\beq
d_{p_{i,\alpha}}\Res_{x(r) \to x(p_{i,\alpha})} { H_{1;0;0}^{(0)}(p_{1,1},q_{1,k_1})f(p_{i,\alpha}) \over
(x(p_{1,1})-x(r))(y(q)-y(r))(x(p_{1,1}) - x(p_{i,\alpha}))H_{1;0;0}^{(0)}(r,q_{1,k_1})}.
\eeq
This residue can be evaluated by using one more time \eq{loopeq} and recalling that only the terms of the form
$H_{0;2;0}(r,p_{i,\alpha})$ have such poles. This finally gives the result, i.e. eq.\ref{soluce}:
\beq\label{soluce2}\encadremath{
\begin{array}{l}
H_{{\bf k_L} ;m;n}^{(g)}({\bf S_L};p_1 ,\dots, p_m;q_1,\dots,q_n) =\cr
\Res_{r \to p_{1,1}, p_{i,\alpha},p_j, \tilde{q}_{1,k_1}^{j}} { H_{1;0;0}^{(0)}(p_{1,1},q_{1,k_1}) \over
(x(p_{1,1})-x(r))(y(q_{1,k_1})-y(r))H_{1;0;0}^{(0)}(r,q_{1,k_1})}\cr
\Big\{
%%%%%%%%%%%%%%%%
%%%%%%%%%%%%%%%%
\sum_{h=1}^{g} H_{0;1;0}^{(h)}(r) H_{k_1,\dots,k_l;m;n}^{(g-h)}(S_1(r),S_2, \dots, S_l;p_1 ,\dots, p_m;q_1,\dots,q_n) \cr
%%%%%%%%%%%%%%%%
%%%%%%%%%%%%%%%%
+\sum_h \sum_{A\bigcup B=\{2, \dots,l\}}  \sum_{I,J} H_{k_1,{\bf k_A};\left|I\right|;\left|J\right|}^{(h)}(S_1(r),{\bf S_A}; {\bf p_{I}};{\bf q_J})
H_{{\bf k_B};m-\left|I\right|+1;n-\left|J\right|}^{(g-h)}({\bf S_B}; r, {\bf p_{M/I}};{\bf q_N/J})\cr
%%%%%%%%%%%%%%%%
%%%%%%%%%%%%%%%%
+ \sum_h \sum_{A\bigcup B=\{2, \dots,l\}}\sum_{\alpha =2}^{k_1} \sum_{I,J} H_{k_1-\alpha+1,{\bf k_B};m-\left|I\right|;n-\left|J\right|}^{(h)}(\{p_{1,\alpha},q_{1,\alpha},\dots p_{1,k_1},q_{1,k_1}\},{\bf S_B}; {\bf p_{M/I}};{\bf q_{N/J}})\cr
\qquad  \times { H_{\alpha-1,{\bf k_A};\left|I\right|;\left|J\right|}^{(g-h)}(\{r,q_{1,1},\dots p_{1,\alpha-1},q_{1,\alpha-1}\},{\bf S_A}; {\bf p_{I}};{\bf q_J})
 \over x(p_{1,\alpha})-x(r)} \cr
%%%%%%%%%%%%%%%%
%%%%%%%%%%%%%%%%
+ \sum_{i=2}^{l} \sum_{\alpha=1}^{k_i} {1 \over x(p_{i,\alpha}) - x(r)}  \times \cr
\quad \times  H_{k_1+k_i,{\bf k_{L/\{1,i\}}};m;n}^{(g)}( \{S_1(r),p_{i,\alpha},q_{i,\alpha},p_{i,\alpha+1}, \dots ,q_{i,k_i},p_{i,1},\dots ,p_{i,\alpha-1},q_{i,\alpha-1}\},{\bf S_{L/\{1,i\}}};{\bf p_M};{\bf q_N})  \cr
%%%%%%%%%%%%%%%%
%%%%%%%%%%%%%%%%
+ \sum_{\alpha =2}^{k_1} { 1 \over x(p_{1,\alpha})-x(r)} \times \cr
 H_{\alpha-1, k_1-\alpha+1, { \bf k_{L/\{1\}}}; m;n}^{(g-1)}(\{r,q_{1,1},\dots p_{1,\alpha-1},q_{1,\alpha-1}\},\{p_{1,\alpha},q_{1,\alpha},\dots p_{1,k_1},q_{1,k_1}\}, {\bf S_{L/\{1\}}};{\bf p_M};{\bf q_N}) \cr
%%%%%%%%%%%%%%%%
%%%%%%%%%%%%%%%%
+ H_{{\bf k_L};m+1;n}^{(g-1)}({\bf S_K}(r);r,{\bf p_M};{\bf q_N})
\Big\} \cr
\end{array}}
\eeq
This recursion equation is a triangular system, thus it allows to compute any $H(S)$ recursively.

\end{document}